\documentclass[12pt]{amsart}
\usepackage{amsmath,amssymb,cite,enumerate,graphicx,hyperref}

\hypersetup{pdftex,colorlinks=true,allcolors=blue}
\usepackage[all]{hypcap}

\usepackage[foot]{amsaddr}
\usepackage[sc]{mathpazo}

\DeclareMathAlphabet\mathsfbi{T1}{phv}{b}{it}

\numberwithin{equation}{section}

\newcommand\BV{\boldsymbol} 
\newcommand\BM{\mathsfbi} 

\newcommand\dif{\mathrm{d}}

\newcommand\parderiv[2]{\frac{\partial #1}{\partial #2}}
\newcommand\deriv[2]{\frac{\mathrm{d} #1}{\mathrm{d} #2}}
\newcommand\EE{\mathbb E}
\newcommand\PP{\mathbb P}
\newcommand\RR{\mathbb R}
\newcommand\cA{\mathcal A}

\newcommand\cL{\mathcal L}
\newcommand\cF{\mathcal F}
\newcommand\cI{\mathcal I}
\newcommand\cH{\mathcal H}
\newcommand\myatop[2]{\genfrac{}{}{0pt}{}{#1}{#2}}

\newcommand\hxinv{\BV q}
\newcommand\resp{R}
\newcommand\jump{J}

\newcommand\pG[2]{p^G_{\BV #1,\BM #2}}

\begin{document}

\title{A theory of average response to large jump perturbations}

\author[Rafail V. Abramov]{Rafail V. Abramov}


\address{Department of Mathematics, Statistics and Computer Science,
University of Illinois at Chicago, 851 S. Morgan st., Chicago, IL 60607}

\email{abramov@uic.edu}

\begin{abstract}
A key feature of the classical Fluctuation Dissipation theorem is its
ability to approximate the average response of a dynamical system to a
sufficiently small external perturbation from an appropriate time
correlation function of the unperturbed dynamics of this system.  In
the present work, we examine the situation where the state of a
nonlinear dynamical system is perturbed by a finitely large,
instantaneous external perturbation (jump) -- for example, the Earth
climate perturbed by an extinction level event.  Such jump can be
either deterministic or stochastic, and in the case of a stochastic
jump its randomness can be spatial, or temporal, or both. We show
that, even for large instantaneous jumps, the average response of the
system can be expressed in the form of a suitable time correlation
function of the corresponding unperturbed dynamics. For stochastic
jumps, we consider two situations: one where a single spatially random
jump of a system state occurs at a predetermined time, and another
where jumps occur randomly in time with small space-time dependent
statistical intensity. For all studied configurations, we compute the
corresponding average response formulas in the form of suitable time
correlation functions of the unperturbed dynamics. Some efficiently
computable approximations are derived for practical modeling
scenarios.
\end{abstract}

\maketitle

\section{Introduction}
\label{sec:intro}

The classical Fluctuation Dissipation theorem (FDT)
\cite{Kub1,Kub2,Ris,Nyq} provides a leading order approximation to the
statistical response of a dynamical system to a small deterministic
external perturbation via statistical correlations of the unperturbed
dynamics. The FDT offered more insight into statistical properties of
dynamical processes near equilibrium in various scientific
applications, such as statistical mechanics of identical particles
\cite{CalGre,EvaMor,KubTodHas}, Ornstein--Uhlenbeck Brownian motion
\cite{Kub1,Kub2,Oka,OrnUhl}, motion of electric charges
\cite{Nyq,VasVli}, turbulence \cite{Kra,Kra2}, quantum field theory
\cite{CalWel,FeyVer,Fuj,Web}, chemical physics \cite{SaiOhm1,SaiOhm2},
and physical chemistry \cite{NoiEzrLor}. In geophysical science, the
FDT was proposed as a sensible approximation for appropriate variables
in a complex climate system \cite{Has,Lei,Lei2} despite the absence of
a classical Gaussian equilibrium state of the traditional statistical
mechanics. This observation spurred a series of works
\cite{Bel,CooHay,CooEslHay,Lei,CarFalIsoPurVul,CohCra,GriBra,Gri,GriBraDym,GriBraMaj,GriDym,KelOrs,MajAbrGer,MajAbrGro,NorBelHar,Pal2},
where various applications of the FDT in the weather and climate
modeling were proposed. In the author's past works
\cite{AbrMaj5,AbrMaj4,AbrMaj6,Abr5,Abr6,Abr7,Abr12,Abr14}, a
computational framework predicting the average response of both
deterministic and stochastic dynamical systems to a small
deterministic or stochastic external perturbation was developed,
studied, and used in a new method for the parameterization of
unresolved processes in reduced models of multiscale dynamics
\cite{Abr9,Abr10,Abr11,AbrKje}.

Thus far, the extensively studied types of external perturbations were
largely limited to small bounded forcing perturbations -- either
deterministic, or in the form of a Brownian motion. However, in many
practical applications the external perturbations are, first, not
necessarily small, and, second, not necessarily in the form of a
bounded external forcing. For example, a physical system may
experience an external ``impulse forcing'' -- that is, the external
forcing in the form of a delta-function, which instantaneously changes
the state of the system. Moreover, this change can be finitely large
-- that is, the delta-function of the impulse forcing may not be
necessarily scaled by a small parameter.

An ubiquitous example of such a global event is known as the
Permian--Triassic extinction \cite{Erw,KnoBamCanGro,Wig}. Due to an
unknown cause, catastrophic global climate changes occurred on Earth
some 250 million years ago between Permian and Triassic periods. As a
result, about 90\% of all living species has become extinct
\cite{HalBig,Par,RenBas, RetSeyKruHolAmb}. One of the accepted
scientific hypotheses is that this event was triggered by a large
meteorite impact \cite{BecPor,BecPor2}.  It is strikingly obvious that
any mathematically and physically adequate explanation or model of
this event cannot possibly assume that this impact was ``small'' in
any reasonable sense -- although one can likely reasonably assume that
this event was ``instantaneous'', at least relative to the time scale
of the subsequent global climate change. In addition, another
reasonable assumption is that such events tend to occur randomly in
time, although they are statistically unlikely to occur frequently.

In the current work, we focus on the scenario where the state of a
nonlinear dynamical system is perturbed by a finitely large,
instantaneous external perturbation, or ``jump''. Such a scenario was
examined previously \cite{BofLacMusVul} for the simple case of a fixed
(that is, independent of the system state) perturbation at a
prescribed time along a single component of the system state
vector. For generality, here we assume that this jump may depend on
the state of the system which immediately precedes the time of the
jump. Furthermore, we assume that the jump can be either deterministic
or stochastic, and for a stochastic jump we further assume that its
randomness can manifest in the spatial configuration of the jump, or
its temporal frequency, or both. We find that the average response of
the system can be expressed in the form of a suitable time correlation
function of the corresponding unperturbed dynamics even for large
instantaneous jumps. In the case of jumps which incorporate a random
component, we examine two scenarios. The first one is where a single
spatially random jump of a system state occurs at a predetermined
time, while the second scenario is the one where jumps occur randomly
in time with small space-time dependent statistical intensity. We also
find that, for all studied configurations, the corresponding average
response formulas are computable in the form of suitable time
correlation functions of the unperturbed dynamics, just like in the
classical FDT formulation. For practical modeling scenarios, we also
derive suitable approximations of the general response formulas which
can be efficiently computed in the context of a numerical simulation.

The work is organized as follows. In Section \ref{sec:unperturbed} we
state the general form of the underlying, unperturbed nonlinear
dynamical system, together with its corresponding forward Kolmogorov
(or Fokker--Planck) equation \cite{App,Oks,Ris}. In Section
\ref{sec:deterministic} we derive the explicit formula of the average
response of the system to a deterministic instantaneous jump
perturbation. In Section \ref{sec:random} we extend the previously
derived average response formula onto the case of a random
instantaneous perturbation, which nonetheless occurs at a prescribed
time. In Section \ref{sec:t_random} we derive the average response
formula for the most general perturbation scenario, where spatially
random jump perturbations also occur at random times. In Section
\ref{sec:summary} we discuss the results and suggest future research
directions.

\section{Unperturbed dynamics}
\label{sec:unperturbed}

In this section, we need to specify the general form of an unperturbed
dynamical system for a typical application in natural sciences. From
what is to be presented below, the basic requirements on the form of
the unperturbed dynamical system can be summarized as follows. First,
the unperturbed dynamical system may incorporate a deterministic
vector field, as well as stochastic effects. We will, however,
restrict the form of the unperturbed dynamical system so that its
solutions are continuous, as it is implied that the discontinuities
will be introduced via the jump perturbations. Additionally, we
require that the unperturbed system has an explicitly formulated
forward Kolmogorov equation \cite{Oks,Ris} with a differentiable
stationary solution. The corresponding perturbed system must also have
a forward Kolmogorov equation, although we are not going to consider
its stationary solutions, if any.

Other than the points listed above, there do not appear to be any
other fundamental constraints which would impose further unmitigable
issues. In particular, the presence of the stochastic component of the
dynamics is not necessarily required for the stationary solution of
the forward Kolmogorov equation to be differentiable -- for example,
if the deterministic vector field preserves a quadratic energy, the
uniform distribution on a constant energy surface is usually both the
stationary statistical state for the system and a smooth distribution
\cite{AbrMaj2,AbrMaj3,AbrKovMaj}.

Thus, throughout the work, we will assume that the unperturbed
dynamical system is described, in general, via the following It\^o
stochastic differential equation:
\begin{equation}
\label{eq:dyn_sys}
\BV x(t)=\BV x_0+\int_0^t\BV f(\BV x(s))\dif s+\int_0^t\BM G(\BV x(s))
\dif\BV W(s).
\end{equation}
Here, $\BV x\in\RR^K$ is the state vector of the system, $\BV W(t)$ is
a $K$-dimensional Wiener process, while $\BV f:\RR^K\to\RR^K$ and $\BM
G:\RR^K\to\RR^{K\times K}$ are smooth vector fields. As motivated
above, the form of \eqref{eq:dyn_sys} is chosen so that it is the most
general form of the L\'evy-type Feller process \cite{Fel2} whose
solutions are almost surely continuous, and which, under a random jump
perturbation, retains the form of the infinitesimal generator
compatible with Courr\`ege's theorem \cite{Cou}, so that the forward
Kolmogorov equation of the jump-perturbed process can be obtained in
an explicit manner \cite{App}.

Let $\psi:\RR^K\to\RR$ be a twice differentiable function with bounded
second derivatives. Let $\EE_{t,t_0}[\psi](\BV x)$ denote the
conditional expectation of $\psi$ at time $t$, provided that the
initial state of the system at time $t_0$ is $\BV x$. Then, the
infinitesimal generator \cite{App,GikSko,Oks} of
$\EE_{t,t_0}[\psi](\BV x)$ is given via:
\begin{equation}
\label{eq:inf_gen}
\left.\parderiv{}t\EE_{t,t_0}[\psi]\right|_{t=t_0}=\cL[\psi]=\BV
f\cdot \parderiv\psi{\BV x}+\frac 12\BM G\BM G^T:
\parderiv{^2\psi}{\BV x^2},
\end{equation}
where ``$:$'' denotes the Frobenius matrix product.

Let $p(t,\BV x)$ denote the probability distribution of solutions of
\eqref{eq:dyn_sys}. Then, we can relate the $p$-average of $\psi$ at
time $t+s$ to the probability density $p(t,\BV x)$ via
\begin{equation}
\label{eq:psi_ts}
\langle\psi\rangle(t+s)=\int_{\RR^K}\psi(\BV x)p(t+s,\BV x)\dif\BV x=
\int_{\RR^K}\EE_{t+s,t}[\psi](\BV x)p(t,\BV x)\dif\BV x.
\end{equation}
The forward partial differential equation for $p(t,\BV x)$ is known
either as the forward Kolmogorov equation \cite{App,GikSko,Oks} or as
the Fokker-Planck equation \cite{Ris}, and is given via
\begin{equation}
\label{eq:kolmogorov}
\parderiv pt+\parderiv{}{\BV x}\cdot (p\BV f)=\frac
12\parderiv{^2}{\BV x^2}:(p\BM G\BM G^T).
\end{equation}
In what follows, we will assume that \eqref{eq:kolmogorov} has a
stationary solution $p_0(\BV x)$,
\begin{equation}
\label{eq:p0}
\parderiv{}{\BV x}\cdot (p_0\BV f)=\frac 12\parderiv{^2}{\BV
  x^2}:(p_0\BM G\BM G^T).
\end{equation}
If the union of all solutions to \eqref{eq:p0} consists of multiple
ergodic components, we will assume below that $p_0(\BV x)$, whose
support we denote as $\cA$, is the indecomposable ergodic component
which is ``physically relevant'' -- that is, a generic initial
condition to \eqref{eq:dyn_sys} almost certainly falls into $\cA$, and
the subsequent jump perturbations are such that the state of the
system in \eqref{eq:dyn_sys} never leaves $\cA$.  This assumption is
necessitated by the need to use the Birkhoff--Khinchin theorem
\cite{Bir,Kol38} for the practical computation of the average response
further below.

As an example of a suitable invariant state, $p_0(\BV x)$ can in its
entirety be ergodic and supported on the whole phase space, which
typically happens when $\BV f$ and $\BM G$ have bounded derivatives of
all orders in $\RR^K$, and, in addition to that, the matrix product
$\BM G\BM G^T$ is uniformly positive definite in $\RR^K$
\cite{Pav}. An elementary example of such a system is the
Ornstein--Uhlenbeck process \cite{OrnUhl}. Besides, the assumption of
ergodicity of $p_0(\BV x)$ is a standard assumption in many works on
this topic
\cite{AbrMaj5,AbrMaj4,AbrMaj6,Abr5,Abr6,Abr7,Abr12,Abr14}. In such a
case, the Birkhoff--Khinchin theorem applies for any finite jump
perturbation of such a system.

Another example is a system whose (possibly deterministic) vector
field preserves the quadratic kinetic energy, in which case $p_0(\BV
x)$ belongs to a family of the corresponding microcanonical Gibbs
states for each constant energy surface
\cite{AbrKovMaj,AbrMaj2,AbrMaj3}.  One can then introduce the jump
perturbation of the state of the system in the form of a ``molecular
collision'' \cite{Abr17,HirCurBir}, given via an explicit formula
which preserves the quadratic energy, which leaves the state of the
system on its assigned constant energy surface after the jump.  More
precisely, let us assume that \eqref{eq:dyn_sys} is deterministic,
possesses the Liouville property
\begin{equation}
\parderiv{}{\BV x}\cdot\BV f(\BV x)=0,
\end{equation}
and preserves the quadratic energy
\begin{equation}
E=\frac 12\|\BV x\|^2.
\end{equation}
along any trajectory $\BV x(t)$. An example of such a system is the
truncated Burgers--Hopf model \cite{AbrKovMaj,MajAbrGro}. It can be
shown \cite{AbrKovMaj,MajAbrGro} that the microcanonical Gibbs state,
which is a uniform probability distribution on the surface of constant
energy, is automatically the invariant statistical state of
\eqref{eq:dyn_sys}.

Let $\BV y, \BV z, \BV n\in\RR^d$ for some positive integer $d\leq
K/2$, with $\|\BV n\|=1$, and let $\BV y'$ and $\BV z'$ be computed
via the transformation
\begin{equation}
\label{eq:yzn}
\BV y'=\BV y+\big((\BV z-\BV y)\cdot\BV n\big)\BV n,\qquad \BV z'=\BV
z+\big((\BV y-\BV z)\cdot\BV n\big)\BV n.
\end{equation}
For $d=3$, the transformation above in \eqref{eq:yzn} describes the
change of velocities during a collision of two hard spheres
\cite{Abr17,HirCurBir}. It can be shown trivially that the sum of
squared norms $\|\BV y\|^2+\|\BV z\|^2$ is preserved by this
transformation. Indeed, observe that
\begin{multline}
\|\BV y'\|^2+\|\BV z'\|^2=\|\BV y+\big((\BV z-\BV y)\cdot\BV n\big)
\BV n\|^2+\|\BV z+\big((\BV y-\BV z)\cdot\BV n\big)\BV n\|^2=\|\BV
y\|^2+\|\BV z\|^2+\\+2\big((\BV z-\BV y)\cdot\BV n\big)\BV n\cdot
\BV y+2\big((\BV y-\BV z)\cdot\BV n\big)\BV n\cdot\BV z+2\big((\BV
z-\BV y)\cdot\BV n\big)^2=\|\BV y\|^2+\|\BV z\|^2.
\end{multline}
Clearly, if $\BV y$ and $\BV z$ are themselves two distinct subsets of
components of $\BV x$, then their replacement with $\BV y'$ and $\BV
z'$ does not change the norm $\|\BV x\|^2$, and, therefore, leaves the
perturbed vector $\BV x$ on its constant energy surface. Thus, the
transformation in \eqref{eq:yzn} is the example of an
energy-preserving jump perturbation of a subset of components of $\BV
x$. Note that, aside from $\|\BV n\|=1$, $\BV n$ is otherwise
arbitrary, and can even be a random variable.

\subsection{Conditional probability density}

In what follows, it is convenient to introduce the conditional
probability density $P_{t,t_0}(\BV x|\BV y)$, defined via
\begin{equation}
\label{eq:cond_P}
\EE_{t,t_0}[\psi](\BV x)=\int_{\RR^K}\psi(\BV y)P_{t,t_0}(\BV y|\BV
x)\dif\BV y.
\end{equation}
From \eqref{eq:psi_ts}, and observing that $\psi$ is arbitrary, it
follows that
\begin{equation}
p(t+s,\BV x)=\int_{\RR^K} P_{t+s,s}(\BV x|\BV y)p(s,\BV y)\dif\BV y.
\end{equation}
Substituting the above equation into \eqref{eq:kolmogorov} and
stripping the integral over $p(\BV y)\dif\BV y$, we obtain the
equation for $P$:
\begin{equation}
\parderiv{}tP_{t,t_0}(\BV x|\BV y)+\parderiv{}{\BV x}\cdot(P_{t,t_0}
(\BV x|\BV y)\BV f)=\frac 12\parderiv{^2}{\BV x^2}:(P_{t,t_0}(\BV x
|\BV y)\BM G\BM G^T),\qquad P_{t_0,t_0}(\BV x|\BV y)=\delta(\BV x-\BV
y).
\end{equation}
Observing that the above equation is autonomous with respect to $t$,
it is clear that
\begin{equation}
P_{t,t_0}(\BV x|\BV y)=P_{t-t_0}(\BV x|\BV y),
\end{equation}
that is, the conditional probability density is a function of
difference between the starting and ending times. Therefore, without
loss of generality, we can assume that $t_0=0$, which yields
\begin{equation}
\label{eq:kolmogorov_P}
\parderiv{}tP_t(\BV x|\BV y)+\parderiv{}{\BV x}\cdot(P_t (\BV x|\BV
y)\BV f)=\frac 12\parderiv{^2}{\BV x^2}:(P_t(\BV x |\BV y)\BM G\BM
G^T),\qquad P_0(\BV x|\BV y)=\delta(\BV x-\BV y).
\end{equation}
Consequently, according to \eqref{eq:cond_P}, the conditional
expectation is also a function of the time difference:
$\EE_{t,t_0}[\psi](\BV x)=\EE_{t-t_0}[\psi](\BV x)$.

\subsection{Special case: the Ornstein--Uhlenbeck process}

Here we consider a special case of the unperturbed dynamical system in
\eqref{eq:dyn_sys}, which has the convenience of being exactly
solvable. Consider the following centered Ornstein--Uhlenbeck process
\cite{OrnUhl}:
\begin{equation}
\label{eq:OU}
\BV x(t)=\BV x_0-\int_0^t\BM L\BV x(s)\dif s+\int_0^t\BM G\dif\BV
W(s).
\end{equation}
Above, $\BM L$ is a constant positive definite $K\times K$ matrix,
while $\BM G$ is a constant $K\times K$ matrix, with $\BM G\BM G^T$
being positive definite. The solution to the Ornstein--Uhlenbeck
process above is given via Duhamel's principle:
\begin{equation}
\BV x(t)=e^{-t\BM L}\BV x_0+\int_0^te^{(s-t)\BM L}\BM G\dif\BV W(s).
\end{equation}
Via the properties of the It\^o integral \cite{Ito,GikSko,Oks}, it can
be readily seen that the expectation of the state variable $\BV x(t)$
satisfies
\begin{equation}
\label{eq:E_x_OU}
\EE_t[\BV x](\BV x_0)=e^{-t\BM L}\BV x_0.
\end{equation}
The stationary probability density $p_0(\BV x)$ of the
Ornstein--Uhlenbeck process in \eqref{eq:OU} is given via
\begin{equation}
\label{eq:p_0_OU}
p_0^{OU}(\BV x)=\frac 1{(2\pi)^{K/2}\sqrt{\det\BM C}}\exp\left(-\frac
12\BV x^T\BM C^{-1}\BV x\right),
\end{equation}
where $\BM C$ is the covariance matrix of the process, given via the
equation
\begin{equation}
\label{eq:cov_OU}
\BM L\BM C+\BM C\BM L^T=\BM G\BM G^T.
\end{equation}
More details on this topic can be found in \cite{Ris}.

\section{A large deterministic jump perturbation}
\label{sec:deterministic}

In this work, the definition of the average response is identical to
that in our earlier works on this topic
\cite{AbrMaj4,AbrMaj5,AbrMaj6,MajAbrGro,Abr5,Abr6,Abr7,Abr12,Abr14}.
Namely, assume that we have a large statistical ensemble of solutions
of \eqref{eq:dyn_sys}. This ensemble is initially distributed
according to $p_0(\BV x)$. At the prescribed time $t_0$, the following
instantaneous jump perturbation is applied to the states of all
ensemble members:
\begin{equation}
\label{eq:det_pert}
\BV x(t_0)=\BV x(t_0-)+\BV h(\BV x(t_0-)),
\end{equation}
Above in \eqref{eq:det_pert}, $\BV h:\RR^K\to\RR^K$ is a continuous
function, and $\BV x(t-)$ denotes the left-limit at $t$. Observe that
the action of $\BV h(\BV x)$ can be interpreted as the instantaneous
jump perturbation of a given trajectory, which, generally, depends on
the pre-jump state (but not on the time of the jump). As noted above,
here we assume that for any $\BV x\in\cA$, $\BV h(\BV x)\in\cA$ --
that is, the state of the system does not leave the support $\cA$ of
the physically relevant ergodic component $p_0(\BV x)$ of
\eqref{eq:dyn_sys} as a result of the jump.

The resulting statistical discrepancy between the perturbed ensemble
and the unperturbed ensemble, as a function of time past the moment of
the perturbation, is what we refer to as the ``average response''. The
key difference between the previous studies and the current work is
that in our previous works a time-dependent forcing was applied at the
initial time and past that, whereas here the perturbed trajectories
continue according to \eqref{eq:dyn_sys} once the jump perturbation
has been applied.

The average response of the statistical ensemble is measured via the
difference between the perturbed and unperturbed conditional
expectations $\EE_t[\psi]$ of a test function $\psi(\BV x)$, averaged
over the equilibrium probability density $p_0(\BV x)$:
\begin{equation}
\label{eq:det_resp}
\Delta\langle\psi\rangle(t-t_0)=\int_{\cA}\big(\EE_{t-t_0}[\psi](\BV x
+\BV h(\BV x))-\EE_{t-t_0}[\psi](\BV x)\big)p_0(\BV x)\dif\BV x,
\end{equation}
where $t_0$ is the time when the external perturbation has occurred.

\subsection{An exactly solvable example}

As an example, we can easily compute the mean state response of the
Ornstein--Uhlenbeck process in \eqref{eq:OU}, with help of
\eqref{eq:E_x_OU}:
\begin{equation}
\label{eq:resp_x_OU}
\Delta\langle\BV x\rangle_{OU}(t-t_0)=e^{(t_0-t)\BM L}\int_{\RR^K}\BV
h(\BV x)p_0^{OU}(\BV x)\dif\BV x,
\end{equation}
where we observe that $p_0^{OU}$ is indecomposable and supported on
the whole space, and thus $\cA$ is the same as $\RR^K$. Above, we only
need to be able to evaluate the integral in terms of elementary
functions, which, given the form of $p_0^{OU}(\BV x)$ in
\eqref{eq:p_0_OU}, leaves a broad choice for $\BV h(\BV x)$ (it can
be, for example, a polynomial in $\BV x$).

\subsection{A practical average response formula for computational
modeling}

It is also a possibility that the dynamical system in
\eqref{eq:dyn_sys} is not exactly solvable, although can be
numerically simulated or modeled. In this case, we need a suitable
adaptation of the response formula in \eqref{eq:det_resp} to the
typical practical restrictions of numerical modeling. Here, we will
assume that, first, there exists an approximation to the stationary
probability distribution $p_0(\BV x)$ in terms of elementary
functions, and, second, there does not exist a similar approximation
to $\EE_t[\psi](\BV x)$. These assumptions are reasonable for a range
of prototype nonlinear dynamical systems, such as the Lorenz 96 model
\cite{MajAbrGro,Lor,LorEma}, the barotropic model of Earth atmosphere
\cite{AbrMaj6,AbrMajKle,Fra}, or the quasigeostrophic 1.5-layer wind
driven double gyre ocean circulation model \cite{AbrMaj7,McC,McCHai}.

In such a situation, let $\hxinv:\RR^K\to\RR^K$ be the inverse of $\BV
x+\BV h(\BV x)$.  Clearly, $\hxinv(\BV x)$ satisfies
\begin{equation}
\label{eq:hxinv_condition}
\hxinv(\BV x)+\BV h(\hxinv(\BV x))=\BV x,\quad \text{for all }\BV x.
\end{equation}
Then, in the perturbed expectation integral, we can change the
variables $\BV x\to\hxinv(\BV x)$ as follows:
\begin{equation}
\int_{\cA}\EE_{t-t_0}[\psi](\BV x+\BV h(\BV x))p_0(\BV x)\dif\BV x=
\int_{\cA}\EE_{t-t_0}[\psi](\BV x)p_0(\hxinv(\BV x))\left|
\parderiv{\hxinv}{\BV x}\right|\dif\BV x,
\end{equation}
where $|\partial\hxinv/\partial\BV x|$ is the Jacobian of $\hxinv(\BV
x)$. The average response formula in \eqref{eq:det_resp} can thus be
written in the form
\begin{multline}
\Delta\langle\psi\rangle(t-t_0)=\int_{\cA}\EE_{t-t_0}[\psi](\BV x)
\left(p_0(\hxinv(\BV x))\left|\parderiv{\hxinv}{\BV x}\right|-p_0(\BV
x)\right)\dif\BV x=\\=\int_{\cA}\EE_{t-t_0}[\psi](\BV x)\left(\frac{
  p_0(\hxinv(\BV x))}{p_0(\BV x)}\left|\parderiv{\hxinv}{\BV x}
\right|-1\right)p_0(\BV x)\dif\BV x,
\end{multline}
where we can divide by $p_0(\BV x)$ since it is nonzero in $\cA$.
Finally, for the purposes of practical computation, we use the
ergodicity property of $p_0(\BV x)$ in $\cA$ and replace, with help of
the Birkhoff--Khinchin theorem \cite{Bir,Kol38}, the measure average
with the following time correlation function over the long-term
trajectory of the unperturbed system in \eqref{eq:dyn_sys}:
\begin{equation}
\label{eq:Dpsi_t}
\Delta\langle\psi\rangle(t-t_0)=\lim_{r\to\infty}\frac 1r\int_0^r\psi(
\BV x(t-t_0+s))\left(\frac{p_0(\hxinv(\BV x(s)))}{p_0(\BV x(s))}\left|
\parderiv{\hxinv}{\BV x}(\BV x(s))\right|-1\right)\dif s,
\end{equation}
where the initial condition for the time series is presumed to be
taken in $\cA$.  The practical computational constraint here (aside
from the necessity for an accurate approximation for $p_0(\BV x)$) is
the ability to invert the function $\BV x+\BV h(\BV x)$. While,
obviously, a large variety of possible forms of the jump function $\BV
h(\BV x)$ is available in general, here we point out a simple form of
$\BV h(\BV x)$ which is explicitly invertible and is likely broad
enough for the majority of practical applications:
\begin{equation}
\label{eq:h}
\BV h(\BV x)=\BV h+\BM H\BV x,
\end{equation}
where $\BV h$ is a constant $K$-vector, and $\BM H$ is a constant
$K\times K$ matrix. For such a form of $\BV h(\BV x)$, we obtain,
explicitly,
\begin{equation}
\label{eq:hxinv}
\hxinv(\BV x)=(\BM I+\BM H)^{-1}(\BV x-\BV h),\qquad\left|\parderiv{
  \hxinv}{\BV x}\right|=\frac 1{|\det(\BM I+\BM H)|}.
\end{equation}
A special case of this scenario was studied previously
\cite{BofLacMusVul} with a fixed perturbation of the form
\begin{equation}
\BV h=(0,\ldots,0,h,0,\ldots,0),
\end{equation}
that is, the fixed jump perturbation was applied to a single component
of the system. It is easy to see that, in such a case, $\hxinv(\BV x)
=\BV x-\BV h$, with its Jacobian being equal to 1.

Finally, we would like to emphasize that none of the average response
formulas above rely on a ``small parameter'' of any kind -- any
imprecisions of the average response computation above will manifest
due to, for example, inaccuracy of the approximation for $p_0(\BV x)$,
or the statistical undersampling of the long-term trajectory of
\eqref{eq:dyn_sys}.

\subsection{The quasi-Gaussian approximation}

The quasi-Gaussian approximation
\cite{AbrMaj4,AbrMaj5,AbrMaj6,MajAbrGro} for the average response
formula in \eqref{eq:Dpsi_t} emerges when the stationary distribution
$p_0(\BV x)$ is replaced by the corresponding Gaussian distribution
with the same mean state and covariance matrix. More precisely, let
$\BV m$ be the mean state of $p_0(\BV x)$, and let $\BM C$ be its
covariance matrix. Then, the quasi-Gaussian approximation for
\eqref{eq:Dpsi_t} is given via
\begin{equation}
\label{eq:Dpsi_t_qG}
\Delta\langle\psi\rangle(t-t_0)=\lim_{r\to\infty}\frac 1r\int_0^r\psi(
\BV x(t-t_0+s))\left(\frac{\pG{m}{C}(\hxinv(\BV x(s)))}{\pG{m}{C}(\BV
  x(s))}\left| \parderiv{\hxinv}{\BV x}(\BV x(s))\right|-1\right)\dif
s,
\end{equation}
where $\pG{m}{C}(\BV x)$ is given explicitly via
\begin{equation}
\label{eq:pG}
\pG{m}{C}(\BV x)=\frac 1{(2\pi)^{K/2}\sqrt{\det\BM C}}\exp\left(-\frac
12 (\BV x-\BV m)^T\BM C^{-1}(\BV x-\BV m)\right).
\end{equation}
The above expression, together with \eqref{eq:hxinv}, renders the
integrand of the time correlation function in
\eqref{eq:Dpsi_t_qG} explicitly computable for a given (computed or
observed) time series $\BV x(t)$, and thus the average response in
\eqref{eq:Dpsi_t_qG} can be computed numerically using standard
methods \cite{AbrMaj4,AbrMaj5,AbrMaj6,MajAbrGro}.

It may seem that, with the replacement of $p_0(\BV x)$ with
$\pG{m}{C}(\BV x)$ in \eqref{eq:Dpsi_t_qG}, one may lose the
information of the relevant ergodic component of $p_0(\BV x)$ (that
is, if the latter is not ergodic on the whole space $\RR^K$). However,
observe that it is not the case -- due to the fact that the time
average in \eqref{eq:Dpsi_t_qG} is computed over the exact time series
of the original dynamical system in \eqref{eq:dyn_sys}, the
statistical sampling of the integrand of \eqref{eq:Dpsi_t_qG} still
occurs on $\cA$ with the distribution $p_0(\BV x)$. Moreover, the most
remarkable property of the time correlation function
in \eqref{eq:Dpsi_t_qG} is that an observed or historically recorded
time series $\BV x(t)$ can be used for its evaluation, without any
need to numerically simulate the dynamical process. This suggests that
the average response could be computed with sufficient degree of
accuracy even for complex geophysical phenomena, as long as the
recorded time series are sufficiently detailed.

\section{A large random jump perturbation}
\label{sec:random}

An interesting generalization of the previous scenario is the
randomization of the jump $\BV h(\BV x)$ (which still occurs at the
time $t_0$). Namely, let $\BV z:\Omega\to\RR^d$ be a random variable
with the distribution measure $\nu$, and let $\BV h(\BV x,\BV
z):\RR^{K+d}\to\RR^K$ be a continuous function of $\BV x$, and bounded
of $\BV z$. Then, it is natural to define the average response of
$\psi$ not only by averaging over the states of the system (that is,
over $p_0(\BV x)$), but also over the possible jumps (that is, over
$\nu$):
\begin{equation}
\label{eq:rand_resp}
\Delta\langle\psi\rangle(t-t_0)=\int_{\cA}\int_{\RR^d}\big(\EE_{t-t_0}[\psi]
(\BV x+\BV h(\BV x,\BV z))-\EE_{t-t_0}[\psi](\BV x)\big)\nu(\dif\BV z)
p_0(\BV x)\dif\BV x.
\end{equation}
Observe that the only difference between \eqref{eq:det_resp} and
\eqref{eq:rand_resp} is the additional average over $\nu$, which
allows to easily extend the average response formulas in
\eqref{eq:resp_x_OU} and \eqref{eq:Dpsi_t} onto the random
perturbation $\BV h(\BV x,\BV z)$ in a straightforward fashion.

\subsection{An exactly solvable example}

The corresponding exactly solvable average response of the mean state
of the Ornstein--Uhlenbeck process is given via
\begin{equation}
\label{eq:rand_resp_x_OU}
\Delta\langle\BV x\rangle_{OU}(t-t_0)=e^{(t_0-t)\BM L}\int_{\RR^K}
\int_{\RR^d}\BV h(\BV x,\BV z)\nu(\dif\BV z)p_0^{OU}(\BV x)\dif\BV x,
\end{equation}
which can be easily seen by taking $\psi(\BV x)=\BV x$ and recalling
the formula for the expectation in \eqref{eq:E_x_OU}. The practical
computability of the mean state response formula in
\eqref{eq:rand_resp_x_OU} now depends, in addition to the choice of
$\BV h(\BV x,\BV z)$, also on the choice of the intensity measure
$\nu$. In the case of $\BV h(\BV x,\BV z)$ being of the form
\begin{equation}
\label{eq:hxz_factored}
\BV h(\BV x,\BV z)=\sum_i\xi_i(\BV z)\BV h_i(\BV x),
\end{equation}
we find that the double integral becomes the sum of the product of
single integrals:
\begin{equation}
\label{eq:rand_resp_x_OU_hz}
\Delta\langle\BV x\rangle_{OU}(t-t_0)=\sum_i e^{(t_0-t)\BM L}\left(
\int_{\RR^d}\xi_i(\BV z)\nu(\dif\BV z)\right)\left(\int_{\RR^K}\BV
h_i(\BV x) p_0^{OU}(\BV x) \dif\BV x\right).
\end{equation}
Here, observe that, for each term in the sum, we obtained the formula
for the mean state response of an Ornstein--Uhlenbeck process to a
deterministic perturbation in \eqref{eq:resp_x_OU}, additionally
multiplied by the integral over $\xi_i(\BV z)\nu(\dif\BV z)$. The
computability of this integral depends entirely on the choice of
$\xi_i(\BV z)$ and $\nu$, and is independent of the rest of the
set-up.

\subsection{A practical average response formula for computational
  modeling}

In order to obtain an analog of the time-averaged response formula in
\eqref{eq:Dpsi_t} for the response to the random perturbation in
\eqref{eq:rand_resp}, we follow the same steps as in the previous
section. We let $\hxinv:\RR^{K+d}\to\RR^K$ be the $\BV x$-inverse of
$\BV x+\BV h(\BV x,\BV z)$ for a given $\BV z$:
\begin{equation}
\label{eq:hxinv_rand_condition}
  \hxinv(\BV x,\BV z)+\BV h(\hxinv(\BV x,\BV z))=\BV x,\quad \text{for
  all }\BV x,\BV z.
\end{equation}
Then, we can again replace $\BV x\to\hxinv(\BV x,\BV z)$ in the
perturbed integral and write
\begin{subequations}
\label{eq:xhx}
\begin{multline}
\int_{\cA}\int_{\RR^d}\EE_{t-t_0}[\psi](\BV x+\BV h(\BV x,\BV z))\nu
(\dif\BV z)p_0(\BV x)\dif\BV x=\\=\int_{\cA}\EE_{t-t_0} [\psi] (\BV
x)\int_{\RR^d} p_0(\hxinv(\BV x,\BV z))\left| \parderiv{\hxinv(\BV
  x,\BV z)}{\BV x}\right|\nu(\dif\BV z)\dif\BV x,
\end{multline}
\begin{equation}
\int_{\cA}\int_{\RR^d}\EE_{t-t_0}[\psi](\BV x)\nu(\dif\BV z) p_0(\BV
x)\dif\BV x=\int_{\cA}\EE_{t-t_0}[\psi](\BV x)p_0(\BV x)\dif\BV x,
\end{equation}
\end{subequations}
where in the unperturbed integral we used the fact that $\nu$ is
normalized to 1, and, as before, $|\partial\hxinv/\partial\BV x|$ is
the Jacobian of $\hxinv(\BV x,\BV z)$ in the $\BV x$-variable. The
average response formula in \eqref{eq:rand_resp} can thus be written
in the form
\begin{equation}
\Delta\langle\psi\rangle(t-t_0)=\int_{\cA}\EE_{t-t_0}[\psi](\BV x)
\left(\frac{\jump(\BV x)}{p_0(\BV x)}-1\right)p_0(\BV x)\dif\BV x,
\end{equation}
where
\begin{equation}
\label{eq:nu_integral}
\jump(\BV x)=\int_{\RR^d}p_0(\hxinv(\BV x,\BV z)) \left| \parderiv{
  \hxinv(\BV x,\BV z)}{\BV x} \right|\nu(\dif\BV z).
\end{equation}
Upon the application of the Birkhoff--Khinchin theorem
\cite{Bir,Kol38}, the average response formula becomes
\begin{equation}
\label{eq:rand_Dpsi_t}
\Delta\langle\psi\rangle(t-t_0)=\lim_{r\to\infty}\frac 1r\int_0^r\psi(\BV
x(t-t_0+s))\left(\frac{\jump(\BV x(s))}{p_0(\BV x(s))}-1\right)\dif s.
\end{equation}
Here, again observe that the average response formula above does not
rely on a small parameter of any kind. Any imprecisions of the average
response computation above will manifest due to, for example,
inaccuracy of the approximation for $p_0(\BV x)$, or a possible
approximation for $\jump(\BV x)$, or the statistical undersampling of
the long-term trajectory of \eqref{eq:dyn_sys}.

However, note that for an efficient numerical computation of the time
correlation function in \eqref{eq:rand_Dpsi_t}, the
term $\jump(\BV x)$ in \eqref{eq:nu_integral} must be expressed in
terms of elementary functions. This requirement, obviously, places
restrictions on the choice of $\nu$, $\BV h(\BV x,\BV z)$, and the
approximation for $p_0(\BV x)$. Below, we consider a few special cases
where the $\nu$-integral above is explicitly computable.

\subsection{Special case 1: a set of spatially pre-determined jumps}

As a simplest example where the $\nu$-integral is explicitly
computable, we consider the following special case. Assume that,
instead of a single deterministic jump, several different, yet
spatially pre-determined jumps may occur randomly with prescribed
probabilities. In other words, the random variable $\BV
z:\Omega\to\RR^d$ may return only a finite set of values $\{\BV
z_j\}$, each with probability $\gamma_j$, $j=1\ldots Q$. In this case,
we can define
\begin{equation}
\label{eq:nu_delta}
\nu(\dif\BV z)=\sum_{j=1}^Q\gamma_j\delta(\BV z-\BV z_j)\dif\BV
z,\qquad\sum_{j=1}^Q\gamma_j=1,
\end{equation}
and, therefore,
\begin{equation}
\jump(\BV x)=\sum_{j=1}^Q\gamma_jp_0(\hxinv(\BV x,\BV z_j)) \left|
\parderiv{\hxinv(\BV x,\BV z_j)}{\BV x}\right|.
\end{equation}
Observe that, as long as $p_0(\BV x)$ and $\hxinv(\BV x,\BV z)$ are
available in the form of explicit formulas, $\jump(\BV x)$ is also an
explicit function of $\BV x$, which allows to compute the time
correlation function in \eqref{eq:rand_Dpsi_t}
efficiently. For a practically computable example of $\BV h(\BV x,\BV
z)$, we can generalize \eqref{eq:h} to include the $\BV z$-dependence:
\begin{equation}
\label{eq:hz}
\BV h(\BV x,\BV z)=\BV h(\BV z)+\BM H(\BV z)\BV x,
\end{equation}
where the $K$-vector $\BV h(\BV z)$ and $K\times K$ matrix $\BM H(\BV
z)$ are known, explicit functions of $\BV z$. This leads to
\begin{equation}
\hxinv(\BV x,\BV z)=(\BM I+\BM H(\BV z))^{-1}(\BV x-\BV h(\BV z)),
\qquad\left| \parderiv{\hxinv(\BV x,\BV z)}{\BV x}\right|=\frac
1{|\det(\BM I+\BM H(\BV z))|},
\end{equation}
and, subsequently,
\begin{equation}
\jump(\BV x)=\sum_{j=1}^Q\gamma_j\frac{p_0\left((\BM I+\BM H(\BV
  z_j))^{-1}(\BV x-\BV h(\BV z_j))\right)}{|\det(\BM I+\BM H(\BV z_j))|}.
\end{equation}

\subsection{Special case 2: a Gaussian jump distribution}

As a practical example, let us consider the scenario where
$\nu(\dif\BV z)$ has a Gaussian density, with its own mean state
vector $\BV m_\nu$ and covariance matrix $\BM C_\nu$:
\begin{equation}
\label{eq:nu_G}
\nu(\dif\BV z)=\pG{m_\nu}{C_\nu}(\BV z)\dif\BV z=\frac 1{(2\pi)^{d/2}
  \sqrt{\det\BM C_\nu}}\exp\left( -\frac 12(\BV z-\BV m_\nu)^T\BM
C_\nu^{-1}(\BV z-\BV m_\nu)\right) \dif\BV z.
\end{equation}
For practical computation, we restrict the $\BV z$-dependent jump
function $\BV h(\BV x,\BV z)$ in \eqref{eq:hz} to
\begin{equation}
\label{eq:h_rand}
\BV h(\BV x,\BV z)=\BV h+\BM H\BV x+\BM H_*\BV z,
\end{equation}
with $\BV h$, $\BM H$ and $\BM H_*$ being the constant $K$-vector,
$K\times K$ matrix and $K\times d$ matrix, respectively. This yields
$\hxinv(\BV x,\BV z)$ in the form
\begin{equation}
\label{eq:hxinv_rand}
\hxinv(\BV x,\BV z)=(\BM I+\BM H)^{-1}(\BV x-\BV h-\BM H_*\BV z),
\qquad\left|\parderiv{ \hxinv}{\BV x}\right|=\frac 1{|\det(\BM I+\BM
  H)|}.
\end{equation}
The form of $\hxinv(\BV x,\BV z)$ in \eqref{eq:hxinv_rand} leads to
the following expression for $\jump(\BV x)$ in \eqref{eq:nu_integral}:
\begin{equation}
\jump(\BV x)=\frac 1{|\det(\BM I+\BM H)|} \int_{\RR^d}p_0\left((\BM
I+\BM H)^{-1}(\BV x-\BV h-\BM H_*\BV z)\right) \nu(\dif\BV z).
\end{equation}
To be able to compute the $\nu$-integral above in the form of an
explicit formula, we will assume that $p_0(\BV x)$ is also given by
its quasi-Gaussian approximation $\pG{m}{C}(\BV x)$ in \eqref{eq:pG}.
This leads to
\begin{equation}
\label{eq:J_G}
\jump(\BV x)=\frac 1{(2\pi)^{K/2}|\det(\BM I+\BM H)|\sqrt{\det\BM
    C\det(\BM C_\nu\BM A)}}\exp\left(\frac 12(\BV b^T(\BV x)\BM
A^{-1}\BV b(\BV x)-c(\BV x))\right),
\end{equation}
where the terms $\BM A$, $\BV b(\BV x)$ and $c(\BV x)$ are given via
\begin{subequations}
\label{eq:Abc}
\begin{equation}
\BM A=((\BM I+\BM H)^{-1}\BM H_*)^T\BM C^{-1}(\BM I+\BM H)^{-1}\BM
H_*+\BM C_\nu^{-1},
\end{equation}
\begin{equation}
\BV b(\BV x)=((\BM I+\BM H)^{-1}\BM H_*)^T\BM C^{-1}(\BV m+(\BM I+\BM
H)^{-1}(\BV h-\BV x))-\BM C_\nu^{-1}\BV m_\nu,
\end{equation}
\begin{equation}
c(\BV x)=(\BV m+(\BM I+\BM H)^{-1}(\BV h-\BV x))^T\BM C^{-1}(\BV m+
(\BM I+\BM H)^{-1}(\BV h-\BV x))+\BV m_\nu^T\BM C_\nu^{-1}\BV m_\nu.
\end{equation}
\end{subequations}
The details of the computation are given in the Appendix
\ref{sec:nu_g_resp}.

Observe that the whole expression in \eqref{eq:J_G} is an explicit
function of $\BV x$ ($\BM A$ is constant, $\BV b$ is linear in $\BV
x$, and $c$ is quadratic in $\BV x$), which means that it can be
evaluated efficiently for the computation of the time
correlation function in \eqref{eq:rand_Dpsi_t}.

\subsection{Special case 3: a linear combination of Gaussian densities}

The natural generalization of the above formula can be easily derived
for a scenario where both $p_0(\BV x)$ and $\nu(\dif\BV z)$ are linear
combinations of Gaussian densities:
\begin{subequations}
\begin{equation}
p_0(\BV x)=\sum_{i=1}^P\beta_i\pG{m_i}{C_i}(\BV x),\qquad\sum_i^P
\beta_i=1,
\end{equation}
\begin{equation}
\nu(\dif\BV z)=\sum_{j=1}^Q\pG{m_{\nu,j}}{C_{\nu,j}}(\BV z) \dif\BV
z,\qquad \sum_j^Q\gamma_j=1,
\end{equation}
\end{subequations}
for some integers $P>0$, $Q>0$.  It is easy to verify that the
corresponding integral $\jump(\BV x)$ in \eqref{eq:nu_integral}
becomes
\begin{multline}
\jump(\BV x)=\frac 1{(2\pi)^{K/2}|\det(\BM I+\BM H)|} \\\sum_{i=1}^P
\sum_{ j=1}^Q\frac{\beta_i\gamma_j}{\sqrt{\det\BM C_i\det(\BM
    C_{\nu,j}\BM A_{ij})}}\exp\left(\frac 12\left(\BV b_{ij}^T(\BV
x)\BM A_{ij}^{-1}\BV b_{ij}(\BV x)-c_{ij}(\BV x)\right)\right),
\end{multline}
with
\begin{subequations}
\begin{equation}
\BM A_{ij}=((\BM I+\BM H)^{-1}\BM H_*)^T\BM C_i^{-1}(\BM I+\BM H)^{-1}\BM
H_*+\BM C_{\nu,j}^{-1},
\end{equation}
\begin{equation}
\BV b_{ij}(\BV x)=((\BM I+\BM H)^{-1}\BM H_*)^T\BM C_i^{-1}(\BV m_i
+(\BM I+\BM H)^{-1}(\BV h-\BV x))-\BM C_{\nu,j}^{-1}\BV m_{\nu,j},
\end{equation}
\begin{equation}
c_{ij}(\BV x)=(\BV m_i+(\BM I+\BM H)^{-1}(\BV h-\BV x))^T\BM C_i^{-1}
(\BV m_i+(\BM I+\BM H)^{-1}(\BV h-\BV x))+\BV m_{\nu,j}^T\BM
C_{\nu,j}^{-1}\BV m_{\nu,j},
\end{equation}
\end{subequations}
where $\BV m_i$ and $\BM C_i$ are the mean state and covariance matrix
of $p^G_i(\BV x)$, respectively. Again, observe that $\BM A_{ij}$ are
constants, while $\BV b_{ij}(\BV x)$ and $c_{ij}(\BV x)$ are explicit
functions of $\BV x$. This means that the measure integral can be
evaluated efficiently in the numerical computation of the time
correlation function in \eqref{eq:rand_Dpsi_t}.

\section{A sequence of large random jump perturbations at random times}
\label{sec:t_random}

An important generalization of the previously studied random jump
process is the extension of the randomness of the perturbation onto
time. Recall that the traditional fluctuation dissipation setting
\cite{GriDym,GriBraDym,Abr5,Abr6,Abr7,Abr12,AbrKje,AbrMaj4,AbrMaj5,AbrMaj6,AbrMaj7,MajAbrGro}
typically consists of a statistical ensemble of solutions of the
unperturbed system in \eqref{eq:dyn_sys}, which are initially
distributed according to the invariant probability measure $p_0(\BV
x)$, given via \eqref{eq:p0}. Then, at the initial time $t_0$, a small
deterministic perturbation is added to each member of this statistical
ensemble. This perturbation may be an explicit function of both time
and the state of that ensemble member. Then, the average response of a
function $\psi(\BV x)$ is the difference of the ensemble average
values $\langle\psi\rangle(t)$ between the perturbed and unperturbed
statistical ensembles, starting at time $t_0$:
\begin{equation}
\label{eq:av_resp}
\Delta\langle\psi\rangle(t,t_0)=\int_\cA\psi(\BV x)\big(p^*(t,\BV x)
-p_0(\BV x)\big)\dif\BV x.
\end{equation}
Above, $p^*(t,\BV x)$ is the solution of the forward Kolmogorov
equation of the corresponding perturbed system, with the initial
condition at $t_0$ given via $p^*(t_0,\BV x)=p_0(\BV x)$.

In the current work, we retain the definition of the average response
as in \eqref{eq:av_resp}, but replace the small deterministic
perturbation of the traditional setting with a new, different
perturbation process, which consists of random jumps of finitely large
magnitude. As follows below, we formulate the random perturbation
process in such a way that its jumps satisfy the following general
properties:
\begin{enumerate}
\item The spacing between the times of jumps is random. However, the
  statistical frequency of the jumps (the so-called ``intensity'') is
  explicitly prescribed, and may conditionally depend, in a specified
  way, on both the time and the pre-jump state of the system. We will
  assume that the intensity of the jump times is small.
\item The jumps themselves are also random, whenever they occur.
  However, their spatial statistical distribution can be explicitly
  prescribed, and may conditionally depend, in a specified way, on the
  pre-jump state of the system. It may not, however, depend on the
  time at the moment of the jump. There will be no restriction on how
  large the jumps can be (except that they must be finite).
\end{enumerate}
As we can see, the properties of the jumps above are general enough to
allow for a wide variety of prototype scenarios in practical
applications. Next, we are going to formulate the requisite jump
perturbation process so that it indeed satisfies the properties
delineated above.

\subsection{The formulation of the random perturbation process}

Recall that the unperturbed system in \eqref{eq:dyn_sys} may include a
Wiener process $\BV W(t)$, which in itself is a random variable. Thus,
we introduce the corresponding probability space $(\Omega,\cF,\PP)$
equipped with a filtration $\{\cF_t\}$, such that for $0\leq t_1\leq
t_2<\infty$, $\cF_{t_1}\subseteq\cF_{t_2}\subseteq\cF$. The Wiener
process $\BV W(t)$ in \eqref{eq:dyn_sys}, as well as the solution $\BV
x(t)$, are chosen to be adapted to $\{\cF_t\}$.

At this point, we proceed with the definition of the random process,
which will perturb the trajectory of the dynamical system in
\eqref{eq:dyn_sys} via instantaneous random jumps. Our goal here is to
formulate the process in such a way so as to have explicit control
over the statistical distribution of jump times, as well as the
spatial distribution of the jumps themselves when they occur. We start
with defining the properties of the statistical distribution of jump
times. Here, we choose the statistical intensity of the times, at
which jumps occur, to be of the form
\begin{equation}
\label{eq:jump_intensity}
\lambda(t)=\alpha\eta(t)g(\BV x(t-)).
\end{equation}
Above, $\eta:\RR\to\RR_{>0}$ is a specified bounded strictly positive
function on a real line, $g:\RR^K\to\RR_{>0}$ is a specified bounded
strictly positive function on $\RR^K$, and $\BV x(t-)$ is the
left-limit at $t$ of the perturbed solution. The small constant
scaling parameter $0<\alpha\ll 1$ signifies that the temporal
intensity of jumps must be small.

As we can see, the intensity of jump times $\lambda$ in
\eqref{eq:jump_intensity} is {\em conditional} -- that is, $\lambda$
is by itself a random variable, adapted to $\{\cF_t\}$. This
definition of the intensity in \eqref{eq:jump_intensity} is
specifically chosen to be very broad, since it allows to specify
explicitly both the deterministic dependence on the time $t$ via
$\eta$, as well as the conditional dependence on the perturbed
solution $\BV x(t)$ via $g$. Of course, one can set $g=1$ if only the
deterministic dependence of the intensity $\lambda$ on time $t$ is
required.

To perform the actual jump perturbations of the state of the dynamical
system in \eqref{eq:dyn_sys}, we introduce a Poisson point process
$\BV n:T\times\Omega\to\RR^d$ with the corresponding jump
  process $\Delta\BV n(t)$:
\begin{equation}
\Delta\BV n(t)=\BV n(t)-\BV n(t-).
\end{equation}
We denote the corresponding Poisson random measure of $\BV n(t)$ by
$N$:
\begin{equation}
N([t_0,t_1],A)=\text{number of values of }\Delta\BV n(t)\in
A\subset\RR^d-\{0\}\text{ for }t_0<t\leq t_1.
\end{equation}
As this Poisson process is a part of the external perturbation to
\eqref{eq:dyn_sys} and thus its properties can be specified however
needed, we choose the intensity measure $\nu$ of $N$, given
  via
\begin{equation}
\nu(A)=\EE N(1,A),
\end{equation}
to be normalized to 1, so that the average intensity of jumps of $\BV
n(t)$ is one jump per unit of time.

In order to adjust the intensity of jumps of the homogeneous Poisson
point process $\BV n(t)$ to the conditional intensity $\lambda$ in
\eqref{eq:jump_intensity}, we introduce the following
time-inhomogeneous point process \cite{DalVer,Papa} $\BV
m:T\times\Omega\to\RR^d$ via
\begin{equation}
\label{eq:mn}
\BV m(t)=\BV n(\tau(t)),
\end{equation}
where the compensator $\tau(t)$ is given via
\begin{equation}
\label{eq:tau}
\tau(t)=\int_0^t\lambda(s)\dif s=\alpha\int_0^t\eta(s)g(\BV x(s-))
\dif s.
\end{equation}
From the perspective of the theory for point processes with
conditional intensities \cite{DalVer,Papa}, what is presented above
should be understood in reverse order -- that is, for a point process
$\BV m(t)$, adapted to $\{\cF_t\}$ and with the conditional intensity
of jump times in \eqref{eq:jump_intensity}, the corresponding process
$\BV n(t)$, defined via \eqref{eq:mn}, is a time-homogeneous Poisson
point process, whose jump times have statistical intensity 1. However,
above we chose a more ``constructive'' order of presentation for
better clarity.

Thus far, we have arranged for the conditional intensity of jump times
$\lambda$ in \eqref{eq:jump_intensity} to be specified as a function
of both time and pre-jump state. However, observe that the actual
jumps of $\BV m(t)$ are identical to those of $\BV n(t)$ -- and, in
particular, their spatial statistical distribution is given via the
intensity measure $\nu$, which is independent of the system state.

To define the conditional dependence of a perturbation on the pre-jump
state of the system, we use the same approach as in the previous
section -- namely, we introduce the jump function $\BV h(\BV x,\BV
z):\RR^{K+d}\to\RR^K$, which depends both on the state of the system
$\BV x(t)$, as well as the random jump, generated by $\BV m(t)$. We
now define the perturbed dynamical system via the following
L\'evy-type Feller process \cite{App,Fel2,Cou}:
\begin{equation}
\label{eq:dyn_sys_pert}
\BV x(t)=\BV x_0+\int_0^t\BV f(\BV x(s))\dif s+\int_0^t\BM G(\BV x(s))
\dif\BV W(s)+\int_0^t\int_{\RR^d-\{0\}}\BV h(\BV x(s-),\BV z)
M(\dif s,\dif\BV z),
\end{equation}
where $M$ is the corresponding Poisson random measure of $\BV
m(t)$. Observe that the properties of random jumps are explicitly
specified via the following quantities:
\begin{enumerate}
\item The statistical intensity of jump times is determined through
  the choice of the functions $\eta(t)$ and $g(\BV x)$ in
  \eqref{eq:jump_intensity}, where the former defines the explicit
  dependence of the intensity on time, while the letter defines the
  conditional dependence of the intensity on the pre-jump system
  state.
\item The spatial distribution of the perturbation jumps is defined
  via the intensity measure $\nu$ of the Poisson point process $\BV
  n(t)$ \eqref{eq:mn} in conjunction with the jump function $\BV h(\BV
  x,\BV z)$. Here, the choice of the measure $\nu$ affects the random
  spread of the jumps, which can be further adjusted by the choice of
  $\BV h(\BV x,\BV z)$, while the latter also allows to specify the
  conditional dependence of the jumps on the pre-jump state of the
  system.
\end{enumerate}
Additionally, observe that there is no provision for the perturbation
jumps to be small -- as mentioned before, it is assumed that the jumps
can be finitely large in magnitude. Instead, we require that the
statistical frequency of jump times is small, for which a small
constant scaling parameter $\alpha$ is provided in
\eqref{eq:jump_intensity}.

\subsection{The infinitesimal generator}

To compute the perturbed probability density $p^*(t,\BV x)$ in the
average response formula \eqref{eq:av_resp}, one first needs to obtain
the infinitesimal generator of the perturbed process in
\eqref{eq:dyn_sys_pert}. When the infinitesimal generator is known,
its integration by parts against the probability density of the system
leads to the forward Kolmogorov equation for $p^*(t,\BV x)$
\cite{MajAbrGro,Abr14,Ris}.

In order to obtain the infinitesimal generator of
\eqref{eq:dyn_sys_pert}, we transform the stochastic jump integral in
\eqref{eq:dyn_sys_pert} into an integral of a time-homogeneous Poisson
process, so that the standard It\^o formula \cite{App} for L\'evy-type
Feller processes \cite{Fel2} could be used. With \eqref{eq:mn}, we can
write the stochastic $M$-integral in the perturbed process
\eqref{eq:dyn_sys_pert} as
\begin{multline}
\int_0^t\int_{\RR^d-\{0\}}\BV h(\BV x(s-),\BV z) M(\dif s,\dif\BV z)=
\sum_{\myatop{0<s\leq t}{\Delta\BV m(s)\neq\BV 0}}\BV h(\BV
x(s-),\Delta\BV m(s))=\\=\sum_{\myatop{0<s\leq t}{\Delta\BV
    n(\tau(s))\neq\BV 0}}\BV h(\BV x(s-),\Delta\BV
n(\tau(s)))=\sum_{\myatop{\tau(0)<s\leq\tau(t)}{\Delta\BV n(s)\neq\BV
    0}}\BV h(\BV x(\tau^{-1}(s)-),\Delta\BV
n(s))=\\=\int_{\tau(0)}^{\tau(t)}\int_{\RR^d-\{0\}}\BV h(\BV
x(\tau^{-1}(s)-),\BV z) N(\dif s,\dif\BV z).
\end{multline}%
Now that the stochastic jump integral in the right-hand side of
\eqref{eq:dyn_sys_pert} has been expressed via a time-homogeneous
Poisson point process, we can obtain the infinitesimal generator of
\eqref{eq:dyn_sys_pert} using the It\^o formula \cite{App}. For a
differentiable test function $\psi:\RR^K\to\RR$ we have the following
It\^o formula for \eqref{eq:dyn_sys_pert}:
\begin{equation}
\psi(\BV x(t+\varepsilon))-\psi(\BV x(t))=\cI[t,t+\varepsilon]
+\int_{\tau(t)}^{\tau(t+\varepsilon)}\int_{\RR^d-\{0\}}r\big(\tau^{-1}(s),\BV
z\big) N(\dif s,\dif\BV z).
\end{equation}
Above, $\cI[t,t+\varepsilon]$ denotes the usual It\^o integral from
$t$ to $t+\varepsilon$ for the drift-diffusion process in
\eqref{eq:dyn_sys}, while $r(t,\BV z)$ denotes the following
expression:
\begin{equation}
r(t,\BV z)=\psi\big(\BV x(t-)+\BV h(\BV x(t-),\BV z)\big)-\psi\big(\BV
x(t-)\big).
\end{equation}
In order to obtain the infinitesimal generator, we need to compute the
expectation of the above integral, conditioned on $\{\cF_t\}$. The
expectation of $\cI[t,t+\varepsilon]$ is, of course, given via the
usual It\^o isometry for drift-diffusion processes. For the stochastic
integral over the Poisson random measure $N$, we observe that $\BV
n(t)$ is time-stationary with independent increments, which leads to
the following expression (keep in mind that $\lambda(t)$ is
$\{\cF_t\}$-adapted):
\begin{multline}
\EE\int_{\tau(t)}^{\tau(t+\varepsilon)}\int_{\RR^d-\{0\}}r\big(\tau^{-1}(s),\BV
z\big)N(\dif s,\dif\BV z)=\\=\EE\int_0^{\tau(t+\varepsilon)-\tau(t)}
\int_{\RR^d-\{0\}}r\big(\tau^{-1}(\tau(t)+s),\BV z\big) N(\dif s,\dif
\BV z)=\\=\int_0^{\varepsilon\lambda(t)} \EE\int_{ \RR^d-\{0\}} r\big(
\tau^{-1}(\tau(t)+s),\BV z\big) N(\dif s,\dif\BV z)+ o(\varepsilon)
=\\=\int_0^{\varepsilon\lambda(t)} \int_{\RR^d-\{0\}}\EE
r\big(\tau^{-1} (\tau(t)+s),\BV z\big)\EE N(\dif s,\dif\BV z)+
o(\varepsilon) =\\=\varepsilon \lambda(t)\int_{\RR^d}r(t,\BV
z)\nu(\dif\BV z)+ o(\varepsilon).
\end{multline}
Recalling \eqref{eq:jump_intensity}, we obtain the following
infinitesimal generator for \eqref{eq:dyn_sys_pert}:
\begin{multline}
\label{eq:inf_gen_pert}
\left.\parderiv{}\varepsilon\EE\psi(\BV x(t+\varepsilon))\right|_{
  \varepsilon=0}=\lim_{\varepsilon\to 0}\frac 1\varepsilon\EE\big[\psi
  (\BV x(t+\varepsilon))-\psi(\BV x(t))\big]=\left.\parderiv{}
\varepsilon\EE\cI[t,t+\varepsilon]\right|_{\varepsilon=0}+\\+
\lim_{\varepsilon\to 0}\frac 1\varepsilon\EE\int_{\tau(t)}^{\tau(t+
  \varepsilon)}\int_{\RR^d-\{0\}}r\big(\tau^{-1}(s),\BV z\big)N(\dif
s, \dif\BV z)=\\= \cL\psi+\alpha\eta(t)g(\BV x)\int_{\RR^d}\big[\psi
  (\BV x+\BV h(\BV x,\BV z))-\psi(\BV x)\big] \nu(\dif\BV z).
\end{multline}
Above, $\BV x=\BV x(t)$, and $\cL$ denotes the infinitesimal generator
of the unperturbed system in \eqref{eq:inf_gen}.

\subsection{The forward Kolmogorov equation}

Observe that the only difference between the unperturbed
\eqref{eq:inf_gen} and perturbed \eqref{eq:inf_gen_pert} infinitesimal
generators is that the latter has an additional term corresponding to
the random jump perturbation. Therefore, it is easy to extend the
unperturbed Kolmogorov equation in \eqref{eq:kolmogorov} onto the
perturbed dynamics, as long as the intensity integral in
\eqref{eq:inf_gen_pert} can be integrated by parts against $p\dif\BV
x$. Indeed, omitting $\alpha\eta(t)$ for convenience (as this factor
is independent of $\BV x$), we have
\begin{multline}
\int_\cA g(\BV x)\bigg(\int_{\RR^d}\big(\psi(\BV x+\BV h(\BV x,\BV z))
-\psi(\BV x)\big)\nu(\dif\BV z)\bigg)p(\BV x)\dif\BV x=\\=\int_\cA
g(\BV x)\bigg(\int_{\RR^d}\psi(\BV x+\BV h(\BV x,\BV z))\nu(\dif\BV z)
\bigg)p(\BV x)\dif\BV x-\int_\cA g(\BV x)\psi(\BV x)p(\BV x)\dif\BV x,
\end{multline}
where in the second term we use the fact that $\nu$ is normalized to
1.

In the second term above, the integral in $\psi\dif\BV x$ can be
readily stripped, however, the first term requires an appropriate
change of variables. With help of $\hxinv(\BV x,\BV z)$, we change the
variables $\BV x\to\hxinv(\BV x,\BV z)$ in the first integral in the
right-hand side above:
\begin{multline}
\int_\cA g(\BV x)\bigg(\int_{\RR^d}\psi(\BV x+\BV h(\BV x,\BV z))
\nu(\dif\BV z)\bigg)p(\BV x)\dif\BV x=\\=\int_\cA\bigg( \int_{\RR^d}
g(\hxinv(\BV x,\BV z))p(\hxinv(\BV x,\BV z))\left| \parderiv{
  \hxinv(\BV x,\BV z)}{\BV x}\right|\nu(\dif\BV z)\bigg) \psi(\BV
x)\dif\BV x.
\end{multline}
Combining the terms, stripping the integrals in $\psi\dif\BV x$, and
multiplying by $\alpha\eta(t)$, we obtain the perturbed Kolmogorov
equation in the form
\begin{multline}
\label{eq:kolmogorov_pert}
\parderiv pt+\parderiv{}{\BV x}\cdot (p\BV f)=\frac 12\parderiv{^2}{
  \BV x^2}:(p\BM G\BM G^T)+\\+\alpha\eta(t)\left(\int_{\RR^d}g(\hxinv(
\BV x,\BV z))p(\hxinv(\BV x,\BV z))\left|\parderiv{\hxinv(\BV x,\BV z)
}{\BV x}\right|\nu(\dif\BV z)-g(\BV x)p(\BV x)\right).
\end{multline}
Despite the fact that the type of the external perturbation here is
completely different from what was studied previously
\cite{Bel,CooHay,CooEslHay,Lei,CarFalIsoPurVul,CohCra,GriBra,Gri,GriBraDym,GriBraMaj,GriDym,KelOrs,MajAbrGer,MajAbrGro,NorBelHar,Pal2},
the forward Kolmogorov equation \eqref{eq:kolmogorov_pert} of the
perturbed process in \eqref{eq:dyn_sys_pert} has a very similar
structure to that of a deterministic \cite{MajAbrGro,Ris} or Brownian
motion \cite{Abr14} perturbation. Namely, observe that there is the
unperturbed part (for which we know the steady solution $p_0(\BV x)$),
and the perturbed part, scaled by a small parameter $\alpha$. Thus, to
recover an approximate solution of \eqref{eq:kolmogorov_pert}, we will
proceed in a standard way, which is to expand the solution in
$\alpha$-power series around $p_0$, and recover the first-order
approximation via Duhamel's principle.

\subsection{The leading order response formula}

We now can look for a solution of \eqref{eq:kolmogorov_pert} around
the stationary solution $p_0(\BV x)$ in the power series form
\begin{equation}
\label{eq:power_series}
p(t,\BV x,\alpha)=p_0(\BV x)+\sum_{n=1}^\infty\alpha^n p_n(t,\BV x),
\end{equation}
which, upon substitution, yields for $p_1$
\begin{equation}
\parderiv{p_1}t+\parderiv{}{\BV x}\cdot({p_1}\BV f)=\frac 12\parderiv{
  ^2}{\BV x^2}:(p_1\BM G\BM G^T)+\eta(t)\left(\jump_g(\BV x)-g(\BV x)
p_0(\BV x)\right),
\end{equation}
where by $\jump_g(\BV x)$ we denote the expression
\begin{equation}
\label{eq:nu_g_integral}
\jump_g(\BV x)=\int_{\RR^d}g(\hxinv(\BV x,\BV z)) p_0(\hxinv(\BV x,
\BV z))\left|\parderiv{\hxinv(\BV x,\BV z)}{\BV x}\right|\nu(\dif\BV
z).
\end{equation}
Assuming that the initial condition is given via $p_1(0,\BV x)=0$, it
is clear that $p_1(t,\BV x)$ can be found via Duhamel's principle:
\begin{equation}
p_1(t,\BV x)=\int_{t_0}^t\dif s\,\eta(s)\int_{\cA}P_{t-s}(\BV x|\BV y)
\left(\jump_g(\BV y)-g(\BV y)p_0(\BV y)\right)\dif\BV y,
\end{equation}
where $P_t(\BV x|\BV y)$ is the conditional distribution of the
unperturbed Kolmogorov equation in \eqref{eq:kolmogorov_P}. Here, the
terms under the $\nu$-integral are known. On the other hand, the
transition probability $P_t(\BV x|\BV y)$ is usually unavailable in
the form of an explicit formula.

However, we can circumvent the direct computation of $P_t(\BV x|\BV
y)$ if the average response of a test function $\psi$ is needed.
Indeed, observe that, according to \eqref{eq:av_resp}, we can express
\begin{equation}
\label{eq:av_resp_1}
\Delta\langle\psi\rangle(t)=\int_{\cA}\psi(\BV x)\left( p(t,\BV
x)-p_0(\BV x)\right)\dif\BV x=\alpha\int_{\cA}\psi(\BV x)p_1(t,\BV
x)\dif\BV x+o(\alpha).
\end{equation}
In turn, the integral of $\psi$ against $p_1$ can be expressed via
\begin{multline}
\int_{\cA}\psi(\BV x) p_1(t,\BV x)\dif \BV x=\int_{\cA}\dif\BV x\,
\psi(\BV x)\int_{t_0}^t\dif s\,\eta(s)\int_{\cA}P_{t-s}(\BV x|\BV y)
\left(\jump_g(\BV y)-g(\BV y)p_0(\BV y)\right)\dif\BV y=\\=\int_{t_0}
^t\dif s\,\eta(s)\int_{\cA}\left(\int_{\RR^K}\psi(\BV y)P_{t-s}(\BV
y|\BV x)\dif\BV y\right)\left(\jump_g(\BV x)-g(\BV x)p_0(\BV x)\right)
\dif\BV x=\\=\int_{t_0}^t\resp_\psi(t-s)\eta(s)\dif s,
\end{multline}
where we denote the {\em average response operator} $\resp_\psi$ via
\begin{multline}
\label{eq:Rpsi}
\resp_\psi(t)=\int_{\cA}\left(\int_{\RR^K}\psi(\BV y)P_t(\BV y|\BV
x)\dif\BV y\right)\left(\jump_g(\BV x)-g(\BV x)p_0(\BV x)\right)\dif
\BV x=\\=\int_{\cA} \EE_t[\psi](\BV x)\left(\jump_g(\BV x)-g(\BV x)
p_0(\BV x)\right)\dif \BV x.
\end{multline}

\subsection{An example of an exactly computable average response operator}

In order to obtain the exact formula for the average response operator
\eqref{eq:Rpsi} of the Ornstein--Uhlenbeck process, we, first, observe
that the part of the integral in \eqref{eq:Rpsi} with $\jump_g(\BV x)$
can be written as
\begin{multline}
\int_{\cA} \EE_t[\psi](\BV x)\jump_g(\BV x)\dif\BV x=\int_{\cA}
\int_{\RR^d}\EE_t[\psi](\BV x)g(\hxinv(\BV x,\BV z))p_0(\hxinv(\BV x,
\BV z))\left|\parderiv{\hxinv(\BV x,\BV z)}{\BV x}\right|\nu(\dif\BV
z) \dif\BV x=\\=\int_{\cA}\left(\int_{\RR^d}\EE_t[\psi](\BV x+\BV
h(\BV x,\BV z))\nu(\dif\BV z)\right)g(\BV x)p_0(\BV x)\dif\BV x,
\end{multline}
which leads to $\resp_\psi$ in the form
\begin{equation}
\label{eq:Rpsi_E}
\resp_\psi(t)=\int_{\cA}\left(\int_{\RR^d}\EE_t[\psi](\BV x+\BV h( \BV
x,\BV z))\nu(\dif\BV z)-\EE_t[\psi](\BV x)\right)g(\BV x)p_0(\BV
x)\dif\BV x.
\end{equation}
Now, for $\psi(\BV x)=\BV x$ and the Ornstein--Uhlenbeck process
\eqref{eq:OU}, we have
\begin{equation}
\resp_{\BV x}(t)=e^{-t\BM L}\int_{\RR^K}\left(\int_{\RR^d}\BV h(\BV x,
\BV z)\nu(\dif\BV z)\right)g(\BV x)p_0^{OU}(\BV x)\dif\BV x.
\end{equation}
For the jump function $\BV h(\BV x,\BV z)$ of the form
in \eqref{eq:hxz_factored}, we have
\begin{equation}
\resp_{\BV x}(t)=\sum_ie^{-t\BM L}\left(\int_{\RR^d}\xi_i(\BV z)\nu(
\dif\BV z)\right)\left(\int_{\RR^K}\BV h_i(\BV x)g(\BV x)p_0^{OU}(\BV
x) \dif\BV x\right).
\end{equation}
Observe that the only difference between \eqref{eq:rand_resp_x_OU_hz}
and the formula above is the presence of $g(\BV x)$ in the integral
over $\dif\BV x$. As the invariant measure $p_0^{OU}$ in
\eqref{eq:p_0_OU} is Gaussian, certain forms of $\BV h_i(\BV x)$ and
$g(\BV x)$ allow to compute the integral explicitly -- for example, if
$\BV h_i(\BV x)$ is a polynomial, and $g(\BV x)$ is a Gaussian
function itself (remember that $g(\BV x)>0$, since it is part of the
statistical intensity of random jumps).

\subsection{A practical computational formula for the average response}

To compute \eqref{eq:Rpsi} in practice, one can employ the
Birkhoff--Khinchin theorem and replace the integral over $p_0(\BV
x)\dif\BV x$ with the long-term average over time series $\BV x(t)$ of
the unperturbed system in \eqref{eq:dyn_sys} in the same way it was
done above in \eqref{eq:xhx}--\eqref{eq:rand_Dpsi_t}. This leads to
the average response operator $\resp_\psi(t)$ in the form of the
following time correlation function:
\begin{equation}
\label{eq:Rpsi_t}
\resp_\psi(t)=\lim_{r\to\infty}\frac 1r\int_0^r\psi(\BV x(t+s)) \left(
\frac{\jump_g(\BV x(s))}{p_0(\BV x(s))}-g(\BV x(s))\right)\dif s.
\end{equation}
Observe that, although the external perturbation here is of a
completely different nature than what was considered traditionally
\cite{Bel,CooHay,CooEslHay,Lei,CarFalIsoPurVul,CohCra,GriBra,Gri,GriBraDym,GriBraMaj,GriDym,KelOrs,MajAbrGer,MajAbrGro,NorBelHar,Pal2},
the form of $\resp_\psi(t)$ is similar to that of the classical linear
response operator \cite{GriDym,MajAbrGro,Ris}. Namely, the response
operator above in \eqref{eq:Rpsi_t} is also a time correlation
function, computed over a long-time series of a solution of the
unperturbed system \eqref{eq:dyn_sys}, except that the response
function $\psi(\BV x)$ is multiplied by a different term under the
correlation integral in \eqref{eq:Rpsi_t}.

For an efficient computation of the time correlation
function above in \eqref{eq:Rpsi_t}, it is necessary for the term
$\jump_g(\BV x)$ to be in the form of an explicit formula. Below we
consider a few special cases where the corresponding integral in
\eqref{eq:nu_g_integral} is explicitly computable.

\subsection{Special case 1: a set of spatially pre-determined jumps}

As a simplest example where the $\nu$-integral in
\eqref{eq:nu_g_integral} is explicitly computable, we consider the
following special case. Assume that the time-homogeneous compound
Poisson process $\BV n(t)$ (and, subsequently, the time-inhomogeneous
compound Poisson process $\BV m(t)$) produce a finite set of jumps --
that is, the jump process $\Delta\BV n(t)$ may assume only a finite
set of values $\{\BV z_j\}$, each with probability $\gamma_j$,
$j=1\ldots Q$. In this case, the intensity measure $\nu$ of $\Delta\BV
n(t)$ is the same as in \eqref{eq:nu_delta}, and, therefore,
\begin{equation}
\jump_g(\BV x)=\sum_{j=1}^Q\gamma_j g(\hxinv(\BV x,\BV z_j))
p_0(\hxinv(\BV x,\BV z_j)) \left| \parderiv{\hxinv(\BV x,\BV z_j)}{\BV
  x}\right|.
\end{equation}
Observe that, as long as $p_0(\BV x)$ and $\hxinv(\BV x,\BV z)$ are
available in the form of explicit formulas, $\jump_g(\BV x)$ is also
an explicit function of $\BV x$, which allows to compute the time
correlation function in \eqref{eq:rand_Dpsi_t}
efficiently. For a practically computable example, we can take $\BV
h(\BV x,\BV z)$ of the form \eqref{eq:hz}, which leads to
\begin{equation}
\jump(\BV x)=\sum_{j=1}^Q\gamma_j g\left((\BM I+\BM H(\BV z_j))^{-1}
(\BV x-\BV h(\BV z_j)\right)\frac{p_0\left((\BM I+\BM H(\BV z_j))^{-1}
  (\BV x-\BV h(\BV z_j))\right)}{|\det(\BM I+\BM H(\BV z_j))|}.
\end{equation}

\subsection{Special case 2: a Gaussian jump distribution}

As another practical example, let us consider the scenario where
$\nu(\dif\BV z)$ has a Gaussian density, with its own mean state
vector $\BV m_\nu$ and covariance matrix $\BM C_\nu$ as above in
\eqref{eq:nu_G}. In addition, we choose the same form for jump
function $\BV h(\BV x,\BV z)$ as above in \eqref{eq:h_rand}, which
yields the same $\hxinv(\BV x,\BV z)$ as in \eqref{eq:hxinv_rand}.
This form of $\nu$ and $\hxinv(\BV x,\BV z)$ leads to the following
expression for $\jump_g(\BV x)$ in \eqref{eq:nu_g_integral}:
\begin{multline}
\jump_g(\BV x)=\frac 1{|\det(\BM I+\BM H)|} \int_{\RR^d} g\left((\BM
I+\BM H)^{-1}(\BV x-\BV h-\BM H_*\BV z)\right)\\p_0\left((\BM I+\BM
H)^{-1}(\BV x-\BV h-\BM H_*\BV z)\right)\pG{m_\nu}{C_\nu}(\BV
z)\dif\BV z.
\end{multline}
Here we will further assume that $p_0(\BV x)$ is also given by the
Gaussian density $\pG{m}{C}(\BV x)$ in \eqref{eq:pG} (that is, we use
the quasi-Gaussian approximation for $p_0$).

At this point, the main difference between the current set-up and the
previously examined scenario in \eqref{eq:J_G} is the additional
presence of the function $g(\BV x)$, whose range must lie above
zero. Clearly, the most simple choice of $g(\BV x)=\text{const}>0$
leads back to \eqref{eq:J_G}. Here, we are going to assume that $g(\BV
x)$ is given via the Gaussian function of the form
\begin{equation}
\label{eq:g}
g(\BV x)=\exp\left(-\frac 12(\BV x-\BV m_g)^T\BM C_g^{-1}(\BV x-\BV
m_g)\right)=(2\pi)^{K/2}\sqrt{\det\BM C_g}\pG{m_g}{C_g}(\BV x),
\end{equation}
with the $K$-vector $\BV m_g$ and symmetric positive definite $K\times
K$ matrix $\BM C_g$ being constant parameters. Observe that such form
of $g(\BV x)$ maximizes the statistical intensity of jumps in the
vicinity of the state $\BV x=\BV m_g$, with the ``width'' of the
intensity bump described via $\BM C_g$. This leads to $\jump_g(\BV x)$
of the form
\begin{equation}
\label{eq:J_GG}
\jump_g(\BV x)=\frac 1{(2\pi)^{K/2}|\det(\BM I+\BM H)|\sqrt{\det\BM
    C\det(\BM C_\nu\BM A_g)}}\exp\left(\frac 12(\BV b_g^T(\BV x)\BM
A_g^{-1}\BV b_g(\BV x)-c_g(\BV x))\right),
\end{equation}
with $\BM A_g$, $\BV b_g(\BV x)$ and $c_g(\BV x)$ given via
\begin{subequations}
\label{eq:Abc_g}
\begin{equation}
\BM A_g=((\BM I+\BM H)^{-1}\BM H_*)^T(\BM C^{-1}+\BM C_g^{-1}) (\BM I
+\BM H)^{-1}\BM H_*+\BM C_\nu^{-1},
\end{equation}
\begin{equation}
\BV b_g(\BV x)=((\BM I+\BM H)^{-1}\BM H_*)^T\left((\BM C^{-1}+\BM
  C_g^{-1})(\BM I+\BM H)^{-1}(\BV h-\BV x)+\BM C^{-1}\BV m +\BM
  C_g^{-1}\BV m_g\right)- \BM C_\nu^{-1}\BV m_\nu,
\end{equation}
\begin{multline}
c_g(\BV x)=((\BM I+\BM H)^{-1}(\BV h-\BV x))^T(\BM C^{-1}+\BM
C_g^{-1})( (\BM I+\BM H)^{-1}(\BV h-\BV x)) +\\+ 2((\BM I+\BM
H)^{-1}(\BV h-\BV x))^T(\BM C^{-1}\BV m+\BM C_g^{-1}\BV m_g)+\BV
m^T\BM C^{-1}\BV m+\BV m_g^T\BM C_g^{-1}\BV m_g+ \BV m_\nu^T\BM
C_\nu^{-1}\BV m_\nu.
\end{multline}
\end{subequations}
The details of the computation are given in the Appendix
\ref{sec:nu_g_resp}.

As before, observe that the whole expression above is an explicit
function of $\BV x$ ($\BM A_g$ is constant, $\BV b_g$ is linear in
$\BV x$, and $c_g$ is quadratic in $\BV x$), which means that it can
be evaluated efficiently for the computation of the time
correlation function in \eqref{eq:Rpsi_t}.

\subsection{Special case 3: a linear combination of Gaussian densities}

The natural generalization of the above formula can be easily derived
for a scenario where $p_0(\BV x)$, $\nu(\dif\BV z)$ and $g(\BV x)$ are
linear combinations of Gaussian densities:
\begin{subequations}
\begin{equation}
p_0(\BV x)=\sum_{i=1}^P\beta_i\pG{m_i}{C_i}(\BV x),\qquad\sum_i^P
\beta_i=1,
\end{equation}
\begin{equation}
\nu(\dif\BV z)=\sum_{j=1}^Q\pG{m_{\nu,j}}{C_{\nu,j}}(\BV z) \dif\BV
z,\qquad \sum_j^Q\gamma_j=1,
\end{equation}
\begin{equation}
g(\BV x)=\sum_{k=1}^T(2\pi)^{K/2}\xi_k\sqrt{\det\BM C_{g,k}}
\pG{m_{g,k}}{C_{g,k}}(\BV x),
\end{equation}
\end{subequations}
for some integers $P>0$, $Q>0$, $T>0$.  It is easy to verify that the
corresponding integral $\jump_g(\BV x)$ in \eqref{eq:nu_g_integral}
becomes
\begin{multline}
\jump(\BV x)=\frac 1{(2\pi)^{K/2}|\det(\BM I+\BM H)|} \\\sum_{i=1}^P
\sum_{j=1}^Q\sum_{k=1}^T\frac{\beta_i\gamma_j\xi_k}{\sqrt{\det\BM
    C_i\det(\BM C_{\nu,j}\BM A_{g,ijk})}}\exp\left(\frac 12\left(\BV
b_{g,ijk}^T(\BV x)\BM A_{g,ijk}^{-1}\BV b_{g,ijk}(\BV x)-c_{g,ijk}(\BV
x)\right)\right),
\end{multline}
with
\begin{subequations}
\begin{equation}
\BM A_{g,ijk}=((\BM I+\BM H)^{-1}\BM H_*)^T(\BM C_i^{-1}+\BM
C_{g,k}^{-1}) (\BM I +\BM H)^{-1}\BM H_*+\BM C_{\nu,j}^{-1},
\end{equation}
\begin{multline}
\BV b_{g,ijk}(\BV x)=((\BM I+\BM H)^{-1}\BM H_*)^T\\\left((\BM C_i^{-1}
+\BM C_{g,k}^{-1})(\BM I+\BM H)^{-1}(\BV h-\BV x)+\BM C_i^{-1}\BV m_i
+\BM C_{g,k}^{-1}\BV m_{g,k}\right)- \BM C_{\nu,j}^{-1}\BV m_{\nu,j},
\end{multline}
\begin{multline}
c_{g,ijk}(\BV x)=((\BM I+\BM H)^{-1}(\BV h-\BV x))^T(\BM C_i^{-1}+\BM
C_{g,k}^{-1})((\BM I+\BM H)^{-1}(\BV h-\BV x)) +\\+ 2((\BM I+\BM
H)^{-1}(\BV h-\BV x))^T(\BM C_i^{-1}\BV m_i+\BM C_{g,k}^{-1}\BV
m_{g,k})+\BV m_i^T\BM C_i^{-1}\BV m_i+\BV m_{g,k}^T\BM C_{g,k}^{-1}\BV
m_{g,k}+ \BV m_{\nu,j}^T\BM C_{\nu,j}^{-1}\BV m_{\nu,j},
\end{multline}
\end{subequations}
where $\BV m_i$ and $\BM C_i$ are the mean state and covariance matrix
of $p^G_i(\BV x)$, respectively. Again, observe that $\BM A_{g,ijk}$
are constants, while $\BV b_{g,ijk}(\BV x)$ and $c_{g,ijk}(\BV x)$ are
explicit functions of $\BV x$. This means that the measure integral
can be evaluated efficiently in the numerical computation of the time
correlation function in \eqref{eq:Rpsi_t}.

\section{Accuracy estimates of the leading order average response}

Recall that, while the average response formulas in in Sections
\ref{sec:deterministic} and \ref{sec:random} are exact, the leading
order average response formula for random-time perturbations,
described in Section \ref{sec:t_random}, is approximate. The accuracy
of the formula in question, specified in
\eqref{eq:av_resp_1}--\eqref{eq:Rpsi}, clearly depends on the
magnitude of the scaling parameter $\alpha$.

Recall that in the conventional setting with deterministic external
perturbations \cite{GriDym,MajAbrGro} it is often intuitively clear
how large the external perturbation can be; for example, if an
external forcing is already present in the dynamical system (say, the
constant forcing in the Lorenz 96 system \cite{Lor,LorEma,AbrMaj4}, or
the vorticity forcing in the barotropic model of the atmosphere
\cite{AbrMaj6,Fra}), then one can reason that the external
perturbation should be much smaller than the already present forcing
\cite{GriBraDym}.

On the other hand, observe that, in the present situation, $\alpha$
does not determine the magnitude of the external perturbation -- in
fact, the latter is presumed to be finitely large. Instead, $\alpha$
regulates the statistical frequency of (potentially large) external
perturbations, and it is thus not immediately clear how small the
scaling parameter $\alpha$ should be chosen to ensure the accuracy of
the leading order average response. In what follows, we address this
question to the extent allowed by the generality of problem setting.

\subsection{Accuracy of the average response operator of the
Ornstein--Uhlenbeck process}

First, we examine a special case where some explicit accuracy
estimates can be made. Here, the unperturbed dynamics are given via
the Ornstein--Uhlenbeck process \eqref{eq:OU}, and the response
function $\psi(\BV x)$ is $\BV x$ itself.
  
The perturbed Ornstein--Uhlenbeck process is given via
\begin{equation}
\BV x_\alpha(t)=\BV x_0-\int_0^t\BM L\BV x_\alpha(s)\dif s+\int_0^t\BM
G\dif\BV W(s)+ \int_0^t\int_{\RR^d-\{0\}}\BV h(\BV x_\alpha(s-),\BV z)
M(\dif s,\dif\BV z).
\end{equation}
Applying the expectation to both sides, we arrive at
\begin{equation}
\EE[\BV x_\alpha(t)]=\BV x_0-\int_0^t\BM L\EE[\BV x_\alpha(s)]\dif\BV
s+\int_0^t \int_{\RR^d}\EE[\BV h(\BV x_\alpha(s),\BV z) M(\dif
  s,\dif\BV z)].
\end{equation}
Here, observe that the expectation in the last term is not conditioned
on $\BV x_\alpha(t)$, but rather on $\BV x_0$. Therefore, in order to
split this expectation into the product of independent expectations of
$\BV h$ and $M$ separately, we must set $g(\BV x)=1$, which renders
$M(\dif s,\dif\BV z)$ independent of $\BV x$. In this case, we obtain
\begin{equation}
\EE[\BV h(\BV x_\alpha(s),\BV z) M(\dif s,\dif\BV z)]=\alpha\EE[\BV
h(\BV x_\alpha(s),\BV z)]\eta(s)\nu(\dif\BV z)\dif s.
\end{equation}
Next, we assume that $\BV h(\BV x,\BV z)$ is of the form
\eqref{eq:h_rand}, which further yields
\begin{equation}
\EE[\BV h(\BV x_\alpha(s),\BV z)]=\BV h+\BM H\EE[\BV x_\alpha(s)]+\BM
H^*\BV z,
\end{equation}
and, subsequently,
\begin{equation}
\int_0^t\int_{\RR^d}\EE[\BV h(\BV x_\alpha(s),\BV z) M(\dif s,\dif\BV
  z)]= \alpha\int_0^t\eta(s)(\BV h+\BM H^*\BV{\bar z}+\BM H\EE[\BV
  x_\alpha(s)])\dif s,
\end{equation}
where we denote
\begin{equation}
\BV{\bar z}=\int_{\RR^d}\BV z\:\!\nu(\dif\BV z).
\end{equation}
We thus arrive at the system of ordinary differential equations
\begin{equation}
\deriv{}t\EE[\BV x_\alpha(t)]=-\BM L\EE[\BV x_\alpha(t)]+\alpha
\eta(t)(\BV h+\BM H^*\BV{\bar z}+\BM H\EE[\BV x_\alpha(t)]),
\qquad\EE[\BV x_\alpha(t_0)]=\BV x_0.
\end{equation}
To obtain the exact solution of this system, we further assume, for
convenience, that $\eta(t)=1$, which, together with the earlier
imposed condition $g(\BV x)=1$, sets the temporal intensity of jumps
to $\alpha$ jumps per unit of time, on average. Subsequently, the
equation above is transformed into
\begin{equation}
\deriv{}t\EE[\BV x_\alpha(t)]=(-\BM L+\alpha\BM H)\EE[\BV x_\alpha(t)]
+\alpha(\BV h+\BM H^*\BV{\bar z}),
\end{equation}
and its solution is now given via Duhamel's principle:
\begin{equation}
\EE[\BV x_\alpha(t)]=e^{t(-\BM L+\alpha\BM H)}\BV x_0+\alpha(-\BM
L+\alpha\BM H)^{-1}\left(e^{t(-\BM L+\alpha\BM H)}-\BM I\right) (\BV
h+\BM H^*\BV{\bar z}).
\end{equation}
The expectation of the corresponding unperturbed solution is, of
course, obtained by setting $\alpha=0$ above:
\begin{equation}
\EE[\BV x(t)]=e^{-t\BM L}\BV x_0,
\end{equation}
with the difference between the two given via
\begin{multline}
\EE[\BV x_\alpha(t)]-\EE[\BV x(t)]=\left(e^{t(-\BM L+\alpha\BM H)}-
e^{-t\BM L}\right)\BV x_0+\\+\alpha(-\BM L+\alpha\BM H)^{-1} \left(
e^{t(-\BM L+\alpha\BM H)}-\BM I\right) (\BV h+\BM H^*\BV{\bar z}).
\end{multline}
The exact average response of the mean state is, therefore, given via
\begin{multline}
\Delta\langle\BV x\rangle(t)=\int_{\RR^K}\left(\EE[\BV x_\alpha(t)]
-\EE[\BV x(t)]\right)p_0^{OU}(\BV x_0)\dif\BV x_0=\\=\alpha(\BM L-
\alpha\BM H)^{-1} \left(\BM I-e^{-t(\BM L-\alpha\BM H)}\right)(\BV h+
\BM H^*\BV{\bar z}),
\end{multline}
where we use the fact that
\begin{equation}
\int_{\RR^K}p_0^{OU}(\BV x)\dif\BV x=1,\qquad\int_{\RR^K}\BV x
p_0^{OU}(\BV x)\dif\BV x=\BV 0.
\end{equation}
The leading order response, obtained via the average response operator
for the same $\BV h(\BV x,\BV z)$, $g(\BV x)=1$ and $\eta(t)=1$, is,
however, given via
\begin{equation}
\Delta\langle\BV x\rangle_R(t)=\alpha\BM L^{-1}\left(\BM I-e^{-t\BM
  L}\right)(\BV h+\BM H^*\BV{\bar z}).
\end{equation}
The immediate qualitative difference here is that the leading order
response, computed via the average response operator, is always
bounded regardless of what $\alpha$ and $\BM H$ are. On the other
hand, the exact average response can become unbounded as $t\to\infty$,
as long as $\alpha$ and $\BM H$ are such that $(\BM L-\alpha\BM H)$
has negative eigenvalues.

Next, assume that $(\BM L-\alpha\BM H)$ is positive definite, so that
the exact average response is also bounded as $t\to\infty$. Then, it
is convenient to regard the difference between the infinite-time
responses as a metric of accuracy, since it no longer involves the
matrix exponential. At the infinite time, with the help of Neumann's
series we have
\begin{multline}
\Delta\langle\BV x\rangle(\infty)-\Delta\langle\BV x\rangle_R(t)
=\alpha\left((\BM L- \alpha\BM H)^{-1}-\BM L^{-1}\right)(\BV h+\BM
H^*\BV{\bar z})=\\=\sum_{k=1}^\infty(\alpha\BM L^{-1}\BM H)^k(\alpha
\BM L^{-1})(\BV h+\BM H^*\BV{\bar z}).
\end{multline}
Assuming that the series converge, and that $\|\BM H\|\sim 1$, it is
clear that the error remains small as long as $\alpha\|\BM L^{-1}\|\ll
1$.

To connect the estimate above with the statistical properties of the
dynamics, observe that
\begin{equation}
\BM L^{-1}=\int_0^\infty e^{-t\BM L}\dif t.
\end{equation}
At the same time, observe that, for the Ornstein--Uhlenbeck process in
\eqref{eq:OU}, the time autocorrelation function of the solution $\BV
x(t)$ is given via the regression theorem \cite{Ris}
\begin{multline}
\langle\BV x(t+s)\BV x^T(s)\rangle=\lim_{r\to\infty}\frac 1r\int_0^r
\BV x(t+s)\BV x^T(s)\dif s=\int_{\RR^K}\EE_t[\BV x](\BV x)\BV x^T
p_0^{OU}(\BV x)\dif\BV x=\\=\int_{\RR^K}\left(e^{-t\BM L}\BV x\right)
\BV x^Tp_0^{OU}(\BV x)\dif\BV x=e^{-t\BM L}\langle\BV x\BV x^T\rangle
=e^{-t\BM L}\BM C,\quad\text{or}\quad e^{-t\BM L}=\langle\BV x(t+s)\BV
x^T(s)\rangle\BM C^{-1},
\end{multline}
where $\BM C$ is the covariance matrix \eqref{eq:cov_OU} of the
Ornstein--Uhlenbeck process. Therefore, for $\BM L^{-1}$ we have
\begin{equation}
\BM L^{-1}=\int_0^\infty e^{-t\BM L}\dif t=\left(\int_0^\infty \langle
\BV x(t+s)\BV x(s)^T\rangle\dif t\right)\BM C^{-1},
\end{equation}
that is, the largest eigenvalue of $\BM L^{-1}$ is the statistical
autocorrelation time $T_{corr}$ of the solution $\BV x(t)$ of the
Ornstein--Uhlenbeck process. Thus, the error of the leading order
average response operator remains small as long as
\begin{equation}
\label{eq:alpha_Tcorr}
\alpha T_{corr}\ll 1.
\end{equation}
Recalling that, for $g(\BV x)=1$ and $\eta(t)=1$, the average time
between the jumps is given via $1/\alpha$, it follows that the average
time between the jumps should be much larger than the
autodecorrelation time of the solution of the unperturbed system.

\subsection{A crude accuracy estimate of the general leading order response}

Here we consider a more general setting with the unspecified response
function $\psi(\BV x)$, and the unperturbed dynamics given via
\eqref{eq:dyn_sys}. For convenience, let us, as above for the
Ornstein--Uhlenbeck process, assume that $\eta(t)=1$, so that the
temporal intensity of jumps is time-independent. Just as in the case
of the Ornstein--Uhlenbeck process, here it is convenient to look at
the response of a function $\psi(\BV x)$ at infinite time. Assuming
that the power series in \eqref{eq:power_series} converge as
$t\to\infty$, for the exact response
$\Delta\langle\psi\rangle(\infty)$ and for the leading order
approximation $\Delta\langle\psi\rangle_R(\infty)$, given via
\eqref{eq:Rpsi}, we have, respectively,
\begin{equation}
\Delta\langle\psi\rangle(\infty)=\sum_{n=1}^\infty\alpha_n\int_\cA\psi(\BV
x)p_n(\infty,\BV x)\dif\BV x,\qquad \Delta\langle\psi\rangle_R(\infty)
=\alpha\int_\cA\psi(\BV x)p_1(\infty,\BV x)\dif\BV x.
\end{equation}
The condition for the accuracy of the leading order response is,
obviously, given via
\begin{equation}
\label{eq:accuracy_condition}
\frac{|\Delta\langle\psi\rangle(\infty)-
  \Delta\langle\psi\rangle_R(\infty)|}{
  |\Delta\langle\psi\rangle_R(\infty)|}\ll 1,
\end{equation}
which means that the remainder of the infinite series, starting with
the second order correction, must be much smaller than the leading
order term itself:
\begin{equation}
\left| \sum_{n=2}^\infty\alpha^{n-1}\int_\cA\psi(\BV x)p_n(\infty,\BV
x)\dif \BV x\right|\ll\left|\int_\cA\psi(\BV x)p_1(\infty,\BV x)\dif\BV
x\right|.
\end{equation}
In order to obtain a somewhat simpler requirement for $\alpha$, let us
presume that there exists a constant $T>0$, such that, for all $n>0$,
the following condition holds:
\begin{equation}
\label{eq:T}
\left|\int_\cA\psi(\BV x)p_{n+1}(\infty,\BV x)\dif\BV x\right|\leq T
\left|\int_\cA\psi(\BV x)p_n(\infty,\BV x)\dif\BV x\right|.
\end{equation}
Then, we can estimate
\begin{multline}
\left|\sum_{n=2}^\infty\alpha^{n-1}\int_\cA\psi(\BV x)p_n(\infty,\BV
x) \dif \BV x\right|\leq\sum_{n=2}^\infty\alpha^{n-1} \left| \int_\cA
\psi(\BV x)p_n(\infty,\BV x)\dif\BV x\right|\leq\\\leq\sum_{n=1}^\infty
(\alpha T)^n \left| \int_\cA \psi(\BV x)p_1(\infty,\BV x)\dif\BV
x\right|.
\end{multline}
Therefore, choosing $\alpha$ so as to impose the condition
\begin{equation}
\sum_{n=1}^\infty(\alpha T)^n=\frac{\alpha T}{1-\alpha T}\ll 1,\;\;
\text{or}\;\;\alpha T\ll\frac 12,
\end{equation}
is sufficient to ensure that \eqref{eq:accuracy_condition} holds
automatically.

It remains to determine the physical meaning of the parameter $T$, for
which we need to formally solve the perturbed Kolmogorov equation in
\eqref{eq:kolmogorov_pert}. For convenience, we rewrite the latter as
\begin{equation}
\label{eq:kolmogorov_pert_1}
\parderiv pt=\cL^\dagger[p]+\alpha\cH^\dagger[p],
\end{equation}
where $\cL$ is the infinitesimal generator of the unperturbed dynamics
in \eqref{eq:inf_gen}, while $\cH$ is the perturbation
component of the perturbed infinitesimal generator in
\eqref{eq:inf_gen_pert}, given, for $\eta(t)=1$, via
\begin{subequations}
\begin{equation}
\cH[\psi](\BV x)=g(\BV x)\int_{\RR^d}\big[\psi(\BV x+\BV h(\BV
  x,\BV z))-\psi(\BV x)\big]\nu(\dif\BV z),
\end{equation}
\begin{equation}
\cH^\dagger[p](\BV x)=\int_{\RR^d}g(\hxinv(\BV x,\BV z))
p(\hxinv(\BV x,\BV z))\left| \parderiv{\hxinv(\BV x,\BV z)}{\BV
  x}\right|\nu(\dif\BV z)-g(\BV x)p(\BV x).
\end{equation}
\end{subequations}
Since the perturbed Kolmogorov equation in
\eqref{eq:kolmogorov_pert_1} no longer depends explicitly on time, we
can set $t_0=0$ without loss of generality. Then, in the notations of
the introduced operators, we have, via Duhamel's principle,
\begin{equation}
p_{n+1}(t,\BV x)=\int_0^t\dif s\int_{\cA}P_{t-s}(\BV x|\BV y)
\cH^\dagger p_n(s,\BV y)\dif\BV y=\int_0^t\dif s\int_{\cA} P_s(\BV
x|\BV y)\cH^\dagger p_n(t-s,\BV y)\dif\BV y.
\end{equation}
Now, observe that, for a solution $p(t,\BV x)$ of the unperturbed
Kolmogorov equation in \eqref{eq:kolmogorov}, we can write
\begin{equation}
p(t+s,\BV x)=\int_\cA P_s(\BV x|\BV y)p(t,\BV y)\dif\BV
y=e^{s\cL^\dagger}p(t,\BV x),
\end{equation}
which, in turn, leads to
\begin{equation}
p_{n+1}(t,\BV x)=\int_0^t\dif s\, e^{s\cL^\dagger}\cH^\dagger
p_n(t-s,\BV x).
\end{equation}
Applying the formula recursively, we arrive at
\begin{equation}
p_n(t,\BV x)=\int_0^t\dif s_1\int_0^{t-s_1}\dif s_2\ldots
\int_0^{t-s_1-\ldots-s_{n-1}}\dif s_n\,e^{s_1\cL^\dagger}\cH^\dagger
e^{s_2\cL^\dagger}\cH^\dagger\ldots e^{s_n\cL^\dagger}
\cH^\dagger p_0(\BV x),
\end{equation}
where the integration occurs over the $n$-dimensional simplex with the
edge length $t$. Sending $t\to\infty$ and assuming that all
$p_n(\infty,\BV x)$ are finite, we separate the integrals with the
help of Fubini's theorem:
\begin{multline}
\label{eq:p_n}
p_n(\infty,\BV x)=\int_0^\infty\dif s_1\int_0^\infty\dif s_2\ldots
\int_0^\infty\dif s_n\,e^{s_1\cL^\dagger}\cH^\dagger
e^{s_2\cL^\dagger}\cH^\dagger\ldots e^{s_n\cL^\dagger}
\cH^\dagger p_0(\BV x)=\\=(-\cL^\dagger)^{-1}\cH^\dagger
p_{n-1}(\infty,\BV x)=\left((-\cL^\dagger)^{-1}\cH^\dagger\right)^2
p_{n-2}(\infty,\BV x)=\ldots=\left((-\cL^\dagger)^{-1}
\cH^\dagger\right)^n p_0(\BV x),
\end{multline}
where we make use of the identity
\begin{equation}
(-\cL^\dagger)^{-1}=\left(\int_0^\infty e^{t\cL^\dagger}\dif t\right).
\end{equation}
Recalling the definition of $T$ in \eqref{eq:T}, we find that, for all
$n>0$,
\begin{equation}
\left|\int_\cA\psi(\BV x)(-\cL^\dagger)^{-1}\cH^\dagger
p_n(\infty,\BV x)\dif\BV x\right|\leq T\left|\int_\cA\psi(\BV x)
p_n(\infty,\BV x)\dif\BV x\right|.
\end{equation}
Now, observe that the expressions above can be written as time
correlation functions
\begin{subequations}
\begin{equation}
\int_\cA\psi(\BV x) p_n(\infty,\BV x)\dif\BV x=\int_\cA\psi(\BV x)
\frac{p_n(\infty,\BV x)}{p_0(\BV x)}p_0(\BV x)\dif\BV
x=\left\langle\psi(\BV x(s))\frac{p_n(\infty,\BV x(s))}{p_0(\BV
  x(s))}\right\rangle,
\end{equation}
\begin{multline}
\int_\cA\psi(\BV x)(-\cL^\dagger)^{-1}\cH^\dagger
p_n(\infty,\BV x)\dif\BV x=\int_0^\infty\dif t\int_\cA\psi(\BV
x)e^{t\cL^\dagger} \cH^\dagger p_n(\infty,\BV x)\dif\BV
x=\\=\int_0^\infty\dif t\int_\cA\psi(\BV x)P_t(\BV x|\BV y)
\cH^\dagger p_n(\infty,\BV y)\dif\BV y\dif\BV
x=\\=\int_0^\infty\dif t\int_\cA\EE_t[\psi](\BV
x)\frac{\cH^\dagger p_n(\infty,\BV x)}{p_0(\BV x)}p_0(\BV
x)\dif\BV x=\int_0^\infty\left\langle\psi(\BV x(t+s)) \frac{
  \cH^\dagger p_n(\infty,\BV x(s))}{p_0(\BV
  x(s))}\right\rangle\dif t,
\end{multline}
\end{subequations}
where, with the help of the Birkhoff--Khinchin theorem, we can compute
the correlation functions via the long-term time averages along a
trajectory of the unperturbed system in \eqref{eq:dyn_sys}:
\begin{subequations}
\label{eq:corr_func}
\begin{equation}
\left\langle\psi(\BV x(s)) \frac{p_n(\infty,\BV x(s))}{p_0(\BV
  x(s))}\right\rangle=\lim_{r\to\infty}\frac 1r\int_0^r \psi(\BV
x(s))\frac{p_n(\infty,\BV x(s))}{ p_0(\BV x(s))}\dif s,
\end{equation}
\begin{equation}
\left\langle\psi(\BV x(t+s)) \frac{ \cH^\dagger p_n(\infty,\BV
  x(s))}{p_0(\BV x(s))}\right\rangle=\lim_{r\to\infty}\frac 1r\int_0^r
\psi(\BV x(t+s))\frac{ \cH^\dagger p_n(\infty,\BV x(s))}{
  p_0(\BV x(s))}\dif s.
\end{equation}
\end{subequations}
Thus, it is clear that $T$ is the supremum of all decorrelation times
for the correlation functions in \eqref{eq:corr_func} across all
$n>0$,
\begin{equation}
T=\max_n\left|\left\langle\psi(\BV x(s)) \frac{p_n(\infty,\BV x(s))}{
  p_0(\BV x(s))}\right\rangle^{-1}\int_0^\infty\left\langle\psi(\BV
x(t+s))\frac{\cH^\dagger p_n(\infty,\BV x(s))}{p_0(\BV x(s))}
\right\rangle\dif t\right|,
\end{equation}
where $p_n(\infty,\BV x)$ is given via \eqref{eq:p_n}.

Clearly, it is impossible to compute $T$ above exactly for a general
dynamical system of the form \eqref{eq:dyn_sys}. However, if the given
system is known to have a ``decorrelation time scale'', in the sense
that all ``typical'' time correlation functions decay roughly on the
same scale, then one can use this time scale as a guidance when
estimating suitable values of the parameter $\alpha$ for practical
computations.

\section{Discussion}
\label{sec:summary}

In the current work, we develop a theory for the average response of a
general nonlinear, possibly stochastic, dynamical system to
instantaneous external jump perturbations of its state. In real-world
physical processes, such perturbations could be a result of an
``impulse forcing'' -- that is, an external forcing in the form of a
delta-function. We consider three distinct scenarios of the jump
perturbation; the first is where a deterministic jump perturbation is
applied at a prescribed time, the second is where a spatially random
jump perturbation is applied at a prescribed time, and the third is
where the jump perturbation is spatially random, and also occurs at
random times (although statistically infrequently).

In all scenarios, we derive the formulas for the average response of a
chosen general test function through suitable time
correlation functions of the unperturbed dynamics.
Throughout the study, we never assume that these jump perturbations
are ``small'' in any reasonable or perceived sense; on the contrary,
these perturbations are presumed to be finitely large in all studied
scenarios. Below we compare and contrast the developed formulas for
the average response to jump perturbations with the response formulas
of the classical Fluctuation Dissipation theorem.

\subsection{Similarities to the classical FDT}

We find that, in the same manner as with the classical Fluctuation
Dissipation theorem, the average response of a chosen test function to
a finitely large jump perturbation can be computed in terms of a
suitable correlation function of the time series of
the unperturbed dynamics -- in particular, no explicit knowledge of
the unperturbed dynamical system is required as long as a sufficiently
detailed long-term historical record of its time series is available.
Thus, in both frameworks, one could theoretically use the observed
time series of the corresponding unperturbed real-world process
directly to compute the average response, potentially reducing the
need in direct numerical simulations of the (likely complex, large
scale and computationally expensive) real-world dynamics.

Also, just like with the classical FDT, a suitable approximation of
the probability density of the invariant statistical state of the
unperturbed dynamics is necessary to compute the time
correlation function. In particular, this results
directly in the analog of the quasi-Gaussian FDT formula
\cite{AbrMaj4,AbrMaj5,AbrMaj6,MajAbrGro} for the jump perturbations.

\subsection{Differences from the classical FDT}

First, recall that the classical Fluctuation Dissipation theorem
furnishes a leading order response formula \cite{Abr14,MajAbrGro,Ris}
to an external perturbation, which, in turn, is accurate only if the
magnitude of the perturbation is small enough. In the response
framework for jump perturbations, developed here, the accuracy of the
average response formulas does not depend on the magnitude of the
perturbation jumps. The only leading order approximation used above is
with respect to the statistical intensity of randomly triggered jumps
-- in order for the response formula to be accurate, the jumps should
occur statistically infrequently.

Second, in context of the classical Fluctuation Dissipation theorem,
the average response is computable for a largely arbitrary additive
external perturbation. In the jump-response framework, the average
response formula is not necessarily available for an arbitrary jump
function -- instead, the jump function must satisfy the inverse
condition in \eqref{eq:hxinv_condition} or
\eqref{eq:hxinv_rand_condition}, and the inverse needs to be known
explicitly. Also, in the case of random jumps, the invariant
probability density of the unperturbed dynamics, the statistical
intensity of external jump perturbations, and the spatial distribution
of the random jumps must all be such that corresponding intensity
integral in the average response formula is computable in terms of
elementary functions. The latter requirement is necessary for the
efficient numerical computation of the time
correlation function for the average response.

At the same time, recall that the average response formula for the
classical FDT involves the gradient of the invariant probability
density of the unperturbed dynamics, which means that any
approximation to the latter must also approximate the gradient
sufficiently well. In contrast, no differentiation of the invariant
probability density is present in any of the jump-response formulas
developed here. Potentially, this may mean that less accurate
approximations for the invariant probability density could be used
without compromising the precision of the predicted response to a
considerable extent.

One of the ``nice'' properties of the classical FDT response formula
is that it is linear with respect to the external forcing -- usually
it is a matrix of certain time correlation functions
multiplying the vector of external perturbations. This makes the
classical FDT particularly suitable for the ``inverse problem'', that
is, the computation of the perturbation vector from the known or
measured response. Observe, however, that all of the average response
formulas, developed above for the external jump perturbations, are
nonlinear with respect to the jump function. As a result, the
computation of the jump perturbation which caused a known response --
for example, the computation of the approximate location and magnitude
of the meteorite impact from the measured climate response -- is
naturally a more challenging and interesting problem.

\subsection{Future research directions}

Given the introductory nature of the present work, here we feel
compelled to point out a few relevant directions of future research.

The basic research directions are those driven by the applicability of
the present framework to real-world situations. Arguably, the most
limiting restriction in the presented theory of the average response
is the need of the jump function to have an explicit inverse (whereas
there is no similar requirement in the classical FDT framework). While
the linear jump function used in the present work is likely a viable
option in a variety of scenarios, there are situations where such a
jump function does not work. As an example, one can consider a
scenario where the jumps must occur between the states of equal energy
in an otherwise energy-conserving system -- clearly, the jump function
must be nonlinear to ensure such a property. Thus, the study and
categorization of possible nonlinear (for example, quadratic) jump
functions with explicit inverses should likely be of high priority.

Another question of importance is the already mentioned ``inverse
problem'', that is, the reconstruction of the perturbation from a
known response. While in the classical FDT framework the inverse
problem is technically trivial due to the linearity of the response
with respect to the external forcing, here we can see that the average
response even to a simplest form of the jump perturbation -- a
constant deterministic jump which occurred at a prescribed time -- is
inherently nonlinear. Thus, recovering the jump function from the
measured response is clearly an important problem in the present
framework.

Unlike the classical FDT framework, where the response is obtained by
taking the time-convolution of the linear response operator with the
external forcing, here observe that, in the scenario with a single
jump perturbation at a deterministic time the average response {\em
  is} the time correlation function itself.
Naturally, one expects the response to such a perturbation to decay in
time, so that the dynamical system eventually returns to is
statistical equilibrium state. However, recall that, unlike the
classical linear response, here the time correlation
function is the {\em exact} average response to the external
perturbation, provided that the invariant probability state is known
exactly. Thus, if the jump-response time correlation
function decays to zero, it means that the system relaxes back toward
its equilibrium state, and vice versa.

If a dynamical system is strongly mixing (in addition to being
ergodic), then the decay of its time correlation
functions is guaranteed, and so is the response to a single jump
perturbation.  However, for an ergodic (and even chaotic) nonlinear
dynamical system, being additionally strongly mixing is not a
necessary requirement.  Moreover, there are empirical examples of
chaotic systems which do not appear to be strongly mixing -- for
example, the Lorenz 96 system \cite{Lor,LorEma,MajAbrGro} in weakly
chaotic regimes does not exhibit noticeable decay of time
correlation functions at linearly unstable
wavenumbers. In such a case, it could be possible that the
corresponding dynamical system never ``settles down'' after a single
jump perturbation.

Thus, the question is: are there real-world scenarios in which the
geophysical dynamics do not rapidly return back to statistical
equilibrium after being subjected to an external jump perturbation? In
particular, could the event, which caused the Permian--Triassic
extinction, be an example of such a perturbation, such that the
occasional planetary climate shifts due to that event continue to
occur to this day? The long-time response of the planetary climate
dynamics to jump perturbations is clearly a very interesting problem
in multiple aspects.

\appendix

\section{Computation of the response for the Gaussian distribution
of jumps}
\label{sec:nu_g_resp}

For \eqref{eq:hxinv_rand}, we have
\begin{multline}
\label{eq:pG_hxinv}
\pG{m}{C}(\hxinv(\BV x,\BV z))=\pG{m}{C}\left((\BM I+\BM H)^{-1}(\BV
x-\BV h-\BM H_*\BV z)\right) =\frac 1{(2\pi)^{K/2}\sqrt{\det\BM
    C}}\\\exp\left(-\frac 12 ((\BM I+\BM H)^{-1}(\BV x-\BV h-\BM
H_*\BV z)-\BV m)^T\BM C^{-1}((\BM I+\BM H)^{-1}(\BV x-\BV h-\BM H_*\BV
z)-\BV m)\right).
\end{multline}
To obtain \eqref{eq:J_G}, we first multiply the above expression by
\eqref{eq:nu_G}. The
resulting form of $\jump(\BV x)$ is given via
\begin{equation}
\jump(\BV x)=\frac 1{|\det(\BM I+\BM H)|} \int_{\RR^d}\pG{m}{C}
\left((\BM I+\BM H)^{-1}(\BV x-\BV h-\BM H_*\BV z)\right)
\pG{m_\nu}{C_\nu}(\BV z)\dif\BV z.
\end{equation}
Multiplying the Gaussian densities under the integral, in the argument
of the resulting exponential we have
\begin{multline}
((\BM I+\BM H)^{-1}(\BV x-\BV h-\BM H_*\BV z)-\BV m)^T\BM C^{-1}((\BM
  I+\BM H)^{-1}(\BV x-\BV h-\BM H_*\BV z)-\BV m)+\\+(\BV z-\BV m_\nu
  )^T \BM C_\nu^{-1}(\BV z-\BV m_\nu)=\BV z^T\BM A\BV z+2\BV b^T\BV z
  +c=\\=(\BV z+\BM A^{-1}\BV b)^T\BM A(\BV z+\BM A^{-1}\BV b)+c-\BV
  b^T\BM A^{-1}\BV b,
\end{multline}
where $\BM A$, $\BV b$ and $c$ are given via \eqref{eq:Abc}.
Observing that the integral
\begin{multline}
\int_{\RR^d}\exp\left(-\frac 12((\BV z+\BM A^{-1}\BV b)^T\BM A(\BV
z+\BM A^{-1}\BV b)+c-\BV b^T\BM A^{-1}\BV b)\right)\dif\BV
z=\\=\exp\left(-\frac 12(c-\BV b^T\BM A^{-1}\BV b)\right)
\int_{\RR^d}\exp\left(-\frac 12(\BV z+\BM A^{-1}\BV b)^T\BM A(\BV
z+\BM A^{-1}\BV b)\right)\dif\BV
z=\\=\frac{(2\pi)^{d/2}}{\sqrt{\det\BM A}}\exp\left(\frac 12(\BV
b^T\BM A^{-1}\BV b-c)\right),
\end{multline}
we arrive at
\begin{multline}
\int_{\RR^d}\pG{m}{C}\left((\BM I+\BM H)^{-1}(\BV x-\BV h-\BM H_*\BV
z) \right)\nu(\dif\BV z)=\\=\frac 1{(2\pi)^{K/2}\sqrt{\det\BM C
    \det(\BM C_\nu\BM A)}}\exp\left(\frac 12(\BV b^T(\BV x)\BM A^{-1}
\BV b(\BV x)-c(\BV x))\right),
\end{multline}
and, subsequently, to \eqref{eq:J_G}.

To obtain \eqref{eq:J_GG}, we first multiply $g(\hxinv(\BV x,\BV z))$
in \eqref{eq:g} by \eqref{eq:nu_G} and \eqref{eq:pG_hxinv}. The
resulting form of $\jump_g(\BV x)$ is given via
\begin{multline}
\jump_g(\BV x)=\frac{(2\pi)^{K/2}\sqrt{\det\BM C_g}}{|\det(\BM I+\BM
  H)|} \int_{\RR^d}\pG{m_g}{C_g}\left((\BM I+\BM H)^{-1}(\BV x-\BV
h-\BM H_*\BV z)\right)\\\pG{m}{C}\left((\BM I+\BM H)^{-1}(\BV x-\BV
h-\BM H_*\BV z)\right)\pG{m_\nu}{C_\nu}(\BV z)\dif\BV z.
\end{multline}
Multiplying the Gaussian densities under the integral, in the argument
of the resulting exponential we have
\begin{multline}
((\BM I+\BM H)^{-1}(\BV x-\BV h-\BM H_*\BV z)-\BV m)^T\BM C^{-1}((\BM
  I+\BM H)^{-1}(\BV x-\BV h-\BM H_*\BV z)-\BV m)+\\+((\BM I+\BM
  H)^{-1}(\BV x-\BV h-\BM H_*\BV z)-\BV m_g)^T\BM C_g^{-1}((\BM I+\BM
  H)^{-1}(\BV x-\BV h-\BM H_*\BV z)-\BV m_g)+\\+(\BV z-\BV m_\nu )^T
  \BM C_\nu^{-1}(\BV z-\BV m_\nu)=\BV z^T\BM A_g\BV z+2\BV b_g^T\BV z
  +c_g=\\=(\BV z+\BM A_g^{-1}\BV b_g)^T\BM A_g(\BV z+\BM A_g^{-1}\BV
  b_g)+c_g-\BV b_g^T\BM A_g^{-1}\BV b_g,
\end{multline}
where $\BM A_g$, $\BV b_g(\BV x)$ and $c_g(\BV x)$ are given via
\eqref{eq:Abc_g}. Observing that the integral
\begin{multline}
\int_{\RR^d}\exp\left(-\frac 12((\BV z+\BM A_g^{-1}\BV b_g)^T\BM
A_g(\BV z+\BM A_g^{-1}\BV b_g)+c_g-\BV b_g^T\BM A_g^{-1}\BV
b_g)\right)\dif\BV z=\\=\exp\left(-\frac 12(c_g-\BV b_g^T\BM
A_g^{-1}\BV b_g)\right) \int_{\RR^d}\exp\left(-\frac 12(\BV z+\BM
A_g^{-1}\BV b_g)^T\BM A_g(\BV z+\BM A_g^{-1}\BV b_g)\right)\dif\BV
z=\\=\frac{(2\pi)^{d/2}}{\sqrt{\det\BM A_g}}\exp\left(\frac 12(\BV
b_g^T\BM A_g^{-1}\BV b_g-c_g)\right),
\end{multline}
we arrive at
\begin{multline}
\int_{\RR^d}\pG{m_g}{C_g}\left((\BM I+\BM H)^{-1}(\BV x-\BV h-\BM
H_*\BV z) \right)\pG{m}{C}\left((\BM I+\BM H)^{-1}(\BV x-\BV h-\BM
H_*\BV z) \right)\pG{m_\nu}{C_\nu}(\BV z)\dif\BV z=\\=\frac
1{(2\pi)^{K/2}\sqrt{\det\BM C_g}}\frac 1{(2\pi)^{K/2}\sqrt{\det\BM
    C\det(\BM C_\nu\BM A_g)}}\exp\left(\frac 12(\BV b_g^T(\BV x)\BM
A_g^{-1} \BV b_g(\BV x)-c_g(\BV x))\right),
\end{multline}
and, subsequently, to \eqref{eq:J_GG}.

\end{document}